\documentclass{JHEP3}
\usepackage{amsmath,amssymb}
\usepackage{graphicx}
\usepackage{cite}
\usepackage{epsfig}
\usepackage{subfigure}
\usepackage{multirow} 

\def\gsim{\ \rlap{\raise 3pt \hbox{$>$}}{\lower 3pt \hbox{$\sim$}}\ }
 \def\lsim{\ \rlap{\raise 3pt \hbox{$<$}}{\lower 3pt \hbox{$\sim$}}\ }

\newcommand{\be}{\begin{equation}}
\newcommand{\ee}{\end{equation}}
\newcommand{\bea}{\begin{eqnarray}}
\newcommand{\eea}{\end{eqnarray}}

\renewcommand{\thefigure}{\Roman{figure}}
\newcommand{\eqn}[1]{eq.~(\ref{#1})}
\newcommand{\Eqn}[1]{Eq.~(\ref{#1})}

\newcommand{\Eqns}[2]{Eqs.~(\ref{#1})-(\ref{#2})}
\title{
Supersymmetric Leptogenesis 
 \vspace{-10pt}
}

\author{Chee Sheng Fong  \\ 
C.N. Yang Institute for Theoretical Physics\\
  State University of New York at Stony Brook\\
  Stony Brook, NY 11794-3840, USA.\\
  E-mail: \email{fong@insti.physics.sunysb.edu}}
\author{M.~C.~Gonzalez-Garcia\\
  Instituci\'o Catalana de Recerca i Estudis Avan\c{c}ats (ICREA), \\
  Departament d'Estructura i Constituents de la Mat\`eria and ICC-UB, \\
  Universitat de Barcelona, 
  Diagonal 647, E-08028 Barcelona, Spain.\\
{\rm and:}  \\ 
C.N. Yang Institute for Theoretical Physics\\
  State University of New York at Stony Brook\\
  Stony Brook, NY 11794-3840, USA.\\
  E-mail: \email{concha@insti.physics.sunysb.edu}}
\author{Enrico Nardi\\
  INFN, Laboratori Nazionali di Frascati,\\
  Via E. Fermi 40, I-00044 Frascati, Italy\\
  E-mail: \email{enrico.nardi@lnf.infn.it}}
 \author{J. Racker\\
  Departament d'Estructura i Constituents de la Mat\`eria 
  and ICC-UB, \\  
  Universitat de Barcelona,
  Diagonal 647, E-08028 Barcelona, Spain.\\
   E-mail: \email{racker@ecm.ub.es}}

 \abstract{We study leptogenesis in the supersymmetric standard model
   plus the seesaw. We identify important qualitative differences that
   characterize supersymmetric leptogenesis with respect to the
   non-supersymmetric case. The lepton number asymmetries in fermions
   and scalars do not equilibrate, and are related via a non-vanishing
   gaugino chemical potential.  Due to the presence of new anomalous
   symmetries, electroweak sphalerons couple to winos and higgsinos,
   and QCD sphalerons couple to gluinos, thus modifying the
   corresponding chemical equilibrium conditions. A new constraint on
   particles chemical potentials corresponding to an exactly conserved
   $R$-charge, that also involves the number density asymmetry of the
   heavy sneutrinos, appears.  These new ingredients determine the
   $3\times 4$ matrices that mix up the density asymmetries of the
   lepton flavours and of the heavy sneutrinos.  We explain why in all
   temperature ranges the particle thermodynamic system is
   characterized by the same number of independent
   quantities. Numerical differences with respect to usual treatment
   remain at the ${\cal O}(1)$ level.}

\keywords{Leptogenesis, Supersymmetry, Neutrino Physics, Beyond
  Standard Model} 

 \preprint{YITP-SB-10-27\\}

\begin{document}

\section{Introduction}

Leptogenesis~\cite{fy,leptoreview} is a theoretical mechanism that can
explain the observed matter-antimatter asymmetry of the Universe.  An
initial lepton asymmetry is generated in the out-of-equilibrium decays
of heavy singlet Majorana neutrinos, and is then partially converted
to a baryon asymmetry by anomalous sphaleron
interactions~\cite{Kuzmin:1985mm}.  Heavy Majorana singlet neutrinos
are also a fundamental ingredient of the seesaw model~\cite{ss} that
explains in an elegant way the suppression of the neutrino mass scale 
with respect to all other particle masses of the Standard Model (SM).

The discovery of neutrino oscillations has promoted leptogenesis to an
utmost attractive scenario to explain the origin of the cosmic baryon
asymmetry.  This is because without any fine-tuning of the seesaw
parameters, a neutrino mass scale naturally compatible with the solar
and atmospheric neutrino mass square differences would be optimal to
yield the correct value of the baryon asymmetry.  The possibility of
explaining two apparently unrelated experimental facts (neutrino
oscillations and the baryon asymmetry) within a single framework has
boosted the interest in leptogenesis studies, leading to important
developments in the field, as for example the inclusion of thermal
corrections~\cite{thermal}, spectator
processes~\cite{spectator1,spectator2}, flavour
effects~\cite{barbieri,endoh,flavour1,flavour2,flavour3}, $CP$
asymmetries in scatterings~\cite{CPscatt}, lepton asymmetries from the
decays of the heavier Majorana neutrinos~\cite{N2,Antusch:2010ms}, and
many more.

In spite of all these advancements, we believe that a proper treatment
of leptogenesis in the supersymmetric case is still lacking.
Supersymmetric leptogenesis constitutes a theoretically appealing
generalization of leptogenesis for the following reason: while the SM
equipped with the seesaw provides the simplest way to realize
leptogenesis, such a framework is plagued by an unpleasant fine-tuning
problem. For a non degenerate spectrum of heavy Majorana neutrinos,
successful leptogenesis requires generically a scale for the singlet
neutrino masses that is much larger than the electroweak (EW)
scale~\cite{di} but at the quantum level the gap between these two
scales becomes unstable.  Low-energy supersymmetry (SUSY) can
naturally stabilize the required hierarchy, and this provides a
sounded motivation for studying leptogenesis in the framework of the
supersymmetrized version of the seesaw mechanism.\footnote{In turn,
  supersymmetric leptogenesis introduces a certain conflict between
  the gravitino bound on the reheat temperature and the thermal
  production of the heavy singlets neutrinos \cite{gravi}. In this
  paper we will not be concerned with the gravitino problem, nor with
  its possible ways out.} Supersymmetric leptogenesis has been studied
in several places, both in dedicated studies~\cite{Plumacher:1997ru}
or in conjunction with SM leptogenesis~\cite{thermal}. However,
several features that are specific of supersymmetry in the high
temperature regime relevant for leptogenesis, in which soft
supersymmetry breaking parameters can be effectively set to zero, have
been overlooked or neglected in these studies.  When the new
ingredients are left out, in spite of the large amount of new
reactions, the differences between SM and supersymmetric leptogenesis
can be resumed by means of simple counting of a few numerical
factors~\cite{leptoreview,Strumia:2006qk,PDBproc}, like for example
the number of relativistic degrees of freedom in the thermal bath, the
number of loop diagrams contributing to the CP asymmetries, the
multiplicities of the final states in the decays of the heavy
neutrinos and sneutrinos.

In this paper we show that, in contrast to the naive picture,
supersymmetric leptogenesis is rich of new and non-trivial features,
and genuinely different from the simpler realization within the SM.  A
first important effect that follows from the requirement that the
supersymmetry breaking scale should not exceed by much the $1\,$TeV
scale, is that above a temperature $T\sim 5\times 10^7\,$GeV the
particle and superparticle leptonic density asymmetries do not
equilibrate. It is then mandatory to account in the Boltzmann
equations for the differences in the number density asymmetries of the
boson and fermion degrees of freedom, that can be given in terms of a
non-vanishing gaugino chemical potential.  A second feature is that
when soft supersymmetry breaking parameters are neglected, additional
anomalous global symmetries that involve both $SU(2)$ and $SU(3)$
fermion representations emerge~\cite{Ibanez:1992aj}.  As a
consequence, the EW and QCD sphaleron equilibrium conditions are
modified with respect to the SM, and this yields a different pattern
of sphaleron induced lepton-flavour
mixing~\cite{barbieri,flavour1,flavour2}. In addition, a new
anomaly-free $R$-symmetry can be defined and the corresponding charge,
being exactly conserved, provides a constraint on the particles
density asymmetries that is not present in the SM.  Finally, a careful
counting of the number of constraining conditions versus the overall
number of particle density asymmetries reveals that four independent
quantities, rather than the three of the SM case, are required to give
a complete description of the various particle asymmetries in the
thermal bath, with the additional quantity corresponding to the number
density asymmetry of the heavy scalar neutrinos. In spite of all these
qualitative differences, numerical corrections with respect to the
case when all new effects are neglected remain at the ${\cal O}(1)$
level.  This is because only spectator processes get affected, while
the overall amount of CP violation driving leptogenesis remains the
same than in previous treatments.

In Section~\ref{sec:conditions} we start with a general description of
the consequences of having different reactions dropping out of
equilibrium as we imagine to raise the thermal bath temperature. We
will next discuss in Section~\ref{sec:SE} the superequilibration
regime, in which leptogenesis would proceed much alike in the SM case,
but that in fact can be realized only at temperatures much below the
lower limit for successful leptogenesis. In Section~\ref{sec:NSE} we
discuss the most relevant theoretical issues, that are
particle-sparticle non-superequilibration, the new global symmetries,
the modified sphaleron conditions, and the relevant conservation laws
that constrain the particle density asymmetries.  We also analyze
specific temperature ranges in which leptogenesis can be successful,
and for each range we present results for the matrices that control
the lepton flavour mixing induced by the fast sphaleron processes and
that, differently from the SM case, now have four columns.  In
Section~\ref{sec:results} we discuss the numerical relevance of our
study with respect to previous treatments. Finally, in
Section~\ref{sec:conclusions} we recap our main results and draw the
conclusions.

\section{Chemical equilibrium conditions and conservation laws}
\label{sec:conditions}

At each specific temperature, particle reactions in the early Universe
can be classified according to their (thermally averaged) rates as:

\begin{enumerate}
\item[I] Much faster than the Universe expansion rate.  In the whole
  temperature range that we will consider, the top-quark Yukawa
  interactions and the $SU(3)\times SU(2)\times U(1)$ gauge
  interactions are examples of reactions of this type.
\item[II] Much slower than the expansion.  For example, at temperature
  $T\gsim (1+\tan^2\beta) 10^{5}\,$GeV the rates of processes that
  involve the electron Yukawa coupling are completely negligible with
  respect to the Universe expansion rate.
\item[III]
 Of the order of the expansion.  This class includes the
  neutrino Yukawa interactions responsible for leptogenesis when the
  temperature is of the order of the heavy Majorana neutrino mass
  $M$.
\end{enumerate}

\noindent

% The first two classes of reactions 

Reactions of type~I enforce specific conditions on the chemical
potentials of the particles in the thermal bath.  Reactions of type II
imply that in the relevant temperature interval, some Lagrangian
parameters can be effectively set to zero.  In most cases (but not in
{\it all} cases) this results in exact global symmetries that
correspond to conservation laws for the corresponding charges. These
conservation laws constrain the asymmetries in the number densities of
the particles that carry the conserved charges.  For particles that
are in chemical equilibrium, there is a statistical correspondence
between the boson ($b$) and fermion ($f$) number density asymmetries
$\Delta n_{b,f}\equiv n_{b,f}- \bar n_{b,f}$ and the corresponding
chemical potentials $\mu_{b,f}$.  In the relativistic limit
$m_{b,f}\ll T$ and at first order in $\mu_{b,f}\ll T$ this
correspondence has the particularly simple form:
\be
\label{eq:Dnmu}
\Delta n_{b}=\frac{g_b}{3}
  T^2\mu_b, \qquad\quad 
\Delta n_{f}=\frac{g_f}{6}T^2\mu_f\,,  
\ee
where $g_{b,f}$ are the number of degrees of freedom of the
corresponding particles.  Therefore, if chemical equilibrium is
enforced for all particles entering a conservation law condition, 
the constraints on number density asymmetries implied by reactions of
type~II can be directly translated into constraints for the particles
chemical potentials. However, when a charge conservation condition
involves particle species that in general remain out of chemical
equilibrium, as is the case for states that participate only in
reactions of type~III, such a correspondence does not exist, and the
constraints must be formulated in terms of number density asymmetries.

Conservation conditions related to gauge symmetries, like hypercharge
conservation (see Eqs.\eqref{eq:Ytot}--\eqref{eq:Ytotmu} below), are
somewhat different from the conservation laws stemming from 
reactions of type~II.
This is because EW symmetry restoration at high temperature occurs
when the order parameter $v$ is dynamically driven to zero, which is
different from the cases when some Lagrangian parameters can be
neglected.  However, for our purposes this difference has no
relevance, and we will treat hypercharge conservation in the same way
than other global conservation laws.

Finally, reactions of type~III violate some symmetries, and thus spoil
the corresponding conservation conditions, without being fast enough
to enforce chemical equilibrium.  To this class belong, by assumption,
all the reactions that involve the neutrinos $N$, the sneutrinos
$\tilde N$ and the anti-sneutrinos $\tilde N^*$.  The evolution of
their number densities and, eventually, of the corresponding
non-conserved charges, must then be tracked by means of Boltzmann
equations.

In the following we will analyze the set of conditions that constrain
the particle number density asymmetries.  Whenever possible we will
express these constraints in terms of particles chemical potentials.
This is the appropriate language for the fast reactions of type~I, and
in most cases can be consistently used also for conditions related to
reactions of type~II.  In principle there are as many chemical
potentials, or more precisely as many density asymmetries as there are
particles in the thermal bath. However, there is also a large set of
conditions enforced by processes in chemical equilibrium and by
conservation laws which ensure that just a few independent quantities
are sufficient to determine all the others. In carrying out our
analysis we imagine to start from relatively low temperatures, that
is, from temperatures well below the lower limit for successful
leptogenesis.  As we raise the temperature, both the expansion rate
and the particle reactions rates vary, and this changes the way
particle reactions are assigned to the previous three classes.  More
specifically, we will only consider what happens when some $B-L$
conserving reaction moves {\it instantaneously} from class I to class
II, since for these reactions the regime in which the rates become of
the order of the expansion (class III) is in general a short-lasting
transient with no relevant effects associated to it. Since we are
interested in leptogenesis, we also assume that, for each temperature
regime, the heavy Majorana neutrino masses and couplings are such that
$B-L$ violating processes always belong to the third
class.~\footnote{Assigning $B-L$ violating processes to class I means
  that they are in thermal equilibrium and no $B-L$ asymmetry can be
  generated.  Assigning them to class II means that $B-L$ is
  conserved, and baryogenesis cannot occur.}  According to the
classification scheme described above, we would expect that at each
temperature the thermodynamic system of the particle soup can be
described in terms of the same number of independent quantities.
  This is because when we move one reaction from the first
class to the second one, we lose one constraint from chemical
equilibrium dynamics, but we gain one constraint from a new
conservation law. While this is in general true, as we will see the
detailed way in which this `switching' of conditions is implemented
has some subtleties, since in some cases a non trivial interplay
between redundant conditions and new global symmetries that are
anomalous, and thus do not correspond to conservation laws, is
involved.

This section is organized in the following way: first we list in
Section~\ref{sec:general} the constraints that hold independently of
assuming a regime in which particles and sparticles chemical
potentials equilibrate (superequilibration (SE) regime) or do not
equilibrate (non-superequilibration (NSE) regime).  In
Section~\ref{sec:SE} we list the constraints that hold only in the SE
regime, and in Section~\ref{sec:NSE} the ones that hold in the
temperature range relevant for supersymmetric leptogenesis, when the
temperature is sufficiently high that particles and sparticles
chemical potentials do not equilibrate.

\medskip

\subsection{General constraints}
\label{sec:general}

The supersymmetric seesaw model is  described by the superpotential
of the supersymmetric SM with the additional terms:
\begin{equation}
W=\frac{1}{2}M_{pq}N^c_{p}N^c_{q}+\lambda_{p\alpha}
N^c_{p}\,\ell_{\alpha} H_u,
\label{eq:superpotential}
\end{equation}
where $p,q=1,2,\dots$ label the heavy singlet states in order of
increasing mass, and $\alpha=e,\mu,\tau$ labels the lepton flavour.
In \Eqn{eq:superpotential} $\ell$, $H_u$ and $N^c$, are respectively
the chiral superfields for the lepton and the up-type Higgs $SU(2)$
doublets and for the heavy $SU(2)$ singlet neutrinos defined according
to usual conventions in terms of their left-handed Weyl spinor
components (for example the $N^c$ supermultiplet has scalar component
$\tilde N^*$ and fermion component $N^c_L$). Finally the $SU(2)$ index
contraction is defined as $\ell_{\alpha}
H_u=\epsilon_{\rho\sigma}\ell_{\alpha}^\rho H_u^\sigma$ with
$\epsilon_{12}=+1$.

Let us start at $T\gsim$ few \,TeV, that is well above the EW or SUSY
breaking scale, but low enough that all the SM and SUSY reactions can
be considered in thermal equilibrium.  We first list in items 1.,
2. and 3. the conditions that hold in the whole temperature range that
we will consider ($M_W \ll T \lsim 10^{14}\,$GeV). Conversely, some of
the Yukawa coupling conditions given in items 4. and 5.  will have to
be dropped as the temperature is increased and the corresponding
reactions go out of equilibrium.  For simplicity of notations, in the
following we denote the chemical potentials with the same notation
that labels the corresponding field: $\phi \equiv \mu_\phi$.

\begin{enumerate}

\item At scales much higher than $M_W$, gauge fields have vanishing
  chemical potential $W=B=g=0$~\cite{ha90}. This also implies that all
  the particles belonging to the same $SU(2)$ or $SU(3)$ multiplets 
  have the same chemical potential. For example
  $\phi(I_3=+\frac{1}{2})=\phi(I_3=-\frac{1}{2})$ for a field $\phi$
  that is a doublet of weak isospin $\vec I$, and similarly for color.

\item Fast gluino, wino and bino interactions relate the 
difference between the chemical potentials of the members of a 
given supermultiplet and  the corresponding gaugino chemical potential. 
Furthermore  $\tilde Q +\tilde g_R \to
  Q$,\ $\tilde Q +\tilde W_R \to Q$,\ $\tilde \ell +\tilde W_R \to \ell
  $,\ $\tilde \ell +\tilde B_R \to \ell $,\ where $\ell,\,Q$
  ($\tilde\ell,\,\tilde Q$) denote the (s)lepton and (s)quarks
  left-handed doublets,  and $\tilde W_R$, $\tilde B_R$ and $\tilde g_R$
are respectively right-handed winos, binos and gluinos, enforce chemical equilibrium conditions which 
  imply that all gauginos have the same chemical potential:
\begin{equation}
-\tilde g = Q-\tilde Q=
-\tilde W= \ell-\tilde \ell=-\tilde B,
\end{equation}
where we have introduced $\tilde W$, $\tilde B$ and $\tilde g$
to denote the chemical potential of the {\it left-handed} gauginos.
It follows that the chemical potentials of the SM particles are related 
to the chemical potential of their respective superpartners as 
\
\begin{eqnarray}
  \label{eq:tQtell}
   \tilde{Q},\tilde \ell &=&    Q,\ell+  \tilde g \\
  \label{eq:HuHd}
   H_{u,d} &=&   \tilde H_{u,d}+  \tilde g \\
  \label{eq:tutdte}
   \tilde u,\tilde d,\tilde e  &=&   u,d,e-  \tilde g. 
\end{eqnarray}
The last relation, in which $u,d,e\equiv u_R,d_R,e_R$ denote the
R-handed $SU(2)$ singlets, follows e.g. from $ \tilde u^c_L= u^c_L+
\tilde g$ for the corresponding $L$-handed fields, together with
$u^c_L=-u_R$ and from the analogous relation for the $SU(2)$ singlet
squarks.  Eqs.\eqref{eq:tQtell}--\eqref{eq:tutdte} allow us to express all the
chemical equilibrium relations in terms of the 18 chemical potentials of
the fermions (SM quarks and leptons, higgsinos and gauginos).

\item Before EW symmetry breaking hypercharge is an exactly conserved
  quantity.  Therefore for the total hypercharge of the Universe we
  have 
  \begin{equation}
    \label{eq:Ytot}
    {\cal Y}_{\rm tot} = \sum_{b} \Delta n_b\, y_b +   \sum_{f} \Delta
    n_f\,y_f ={\rm const}, 
  \end{equation}
  where $y_{b,f}$ denotes the hypercharge of the $b$-bosons or
  $f$-fermions.  Given that leptogenesis aims to explain the origin of
  the matter-antimatter asymmetry, it is reasonable to assume as
  initial condition that at sufficiently high temperatures all
  particle-antiparticle density asymmetries vanish $\Delta
  n_{b,f}\big|_{T\to \infty}=0$ which implies $ {\cal Y}_{\rm tot}
  =0$.  By using \eqref{eq:Dnmu} the hypercharge conservation condition
  can be rewritten as:
 \begin{equation}
    \label{eq:Ytotmu}
\!\!\! \frac{1}{3}\, \sum_{b} \mu_b\,g_b\, y_b +  
\frac{1}{6} \sum_{f} \mu_f\,g_f\,y_f =
\sum_i\left(Q_i+2u_i-d_i\right)
-\sum_\alpha\left(\ell_\alpha+e_\alpha\right)+\tilde{H_u}-\tilde{H_d}=
0, 
\end{equation}
where $Q_i$ denote the three quark doublets, $u_i=u,\,c,\,t$ and 
$d_i=d,\,s,\,b$.
In writing down Eq.\eqref{eq:Ytotmu} we
have used Eqs.\eqref{eq:tQtell}--\eqref{eq:tutdte} to express the chemical
potentials of the scalars in terms of those of their fermion
partners. Given that Eqs.\eqref{eq:tQtell}--\eqref{eq:tutdte} 
hold in the whole
temperature range that we will analyze,  the same is true also for
Eq.\eqref{eq:Ytotmu}.

\end{enumerate}

Note that the hypercharge neutrality condition involves only the 17
chemical potentials of the fermionic components of the matter
superfields since the gaugino chemical potential, that is involved in
the substitutions~\eqref{eq:tQtell}-\eqref{eq:tutdte}, eventually cancels
out. This is expected, since all gauginos have vanishing hypercharge;
however, in counting the number of chemical potentials $\tilde g$ must
be included, yielding a total number of 18.

\begin{enumerate}

\item[4.] When the reactions mediated by the leptons Yukawa couplings are
  faster than the Universe expansion rate
$H \sim 1.66 \sqrt{g_*} T^2/M_P$ (where $M_P$ is the Planck mass
  and $g_*=228.75$ in the MSSM), the following chemical
  equilibrium conditions are enforced:
\be 
\label{eq:leptons}
\ell_\alpha - e_\alpha + \tilde H_d + \tilde g =0, \qquad
(\alpha=e,\,\mu,\,\tau).  
\ee
For $\alpha=e$ the corresponding Yukawa condition holds only as long
as
\be
\label{eq:Te}
T\lsim 10^5(1+\tan^2\beta)\,{\rm GeV,} 
\ee
when Yukawa reactions between the first generation left-handed $SU(2)$
lepton doublets $\ell_e$ and the right-handed singlets $e$ are faster
than the expansion~\cite{eR-equilibrium}.  Note also that, as is
discussed in refs.~\cite{AristizabalSierra:2009mq,Fong:2010zu}, if the
temperature is not too low lepton flavour equilibration induced by
off-diagonal slepton soft masses will not occur. We assume that this
is the case, and thus we take the three $\ell_\alpha$ to be 
independent quantities.

\item[5.]  Reactions mediated by the quarks Yukawa couplings enforce the
  following six  chemical equilibrium conditions:
\bea 
\label{eq:upquarks}
Q_i - u_i + \tilde H_u + \tilde g &=&0, \qquad
(u_i=u,\,c,\,t),\\  
\label{eq:downquarks}
Q_i - d_i + \tilde H_d + \tilde g &=&0, \qquad
(d_i=d,\,s,\,b)\,.  
\eea
The up-quark Yukawa coupling maintains chemical equilibrium between
the left and right handed up-type quarks up to $T\sim 2\cdot
10^6\,$GeV.  Note that when the Yukawa reactions of at least two
families of quarks are in equilibrium, the mass basis is fixed for all
the quarks and squarks.  Intergeneration mixing then implies that
family-changing charged-current transitions are also in equilibrium:
$b_L \to c_L$ and $t_L \to s_L$ imply $Q_2 = Q_3$; $s_L \to u_L$ and
$c_L \to d_L$ imply $Q_1 = Q_2$. Thus, up to temperatures $T\lsim
10^{11}\,$GeV, that are of the order of the charm Yukawa coupling
equilibration temperature, the three quark doublets have the same
chemical potential:
  \begin{equation}
    \label{eq:Q}
    Q\equiv Q_3=Q_2=Q_1. 
  \end{equation}
  At higher temperatures, when only the third family is in
  equilibrium, we will have instead $Q\equiv Q_3=Q_2\neq Q_1$.  Above
  $T\sim 10^{13}$ when (for moderate values of $\tan\beta$) also the
  $\tau$ and $b$-quark $SU(2)$ singlets decouple from their Yukawa
  reactions, all intergeneration mixings are also negligible and
  $Q_3\neq Q_2\neq Q_1$.

\end{enumerate}

\medskip

\subsection{Superequilibration regime}
\label{sec:SE}

At relatively low temperatures, additional conditions from reactions
in chemical equilibrium hold. Since the constraints below apply
only in the SE regime, we number them including this label.
\begin{enumerate}
\item[$6_{\rm SE}$.]

  Equilibration of the particle-sparticle chemical potentials
  $\mu_\phi=\mu_{\tilde \phi}$ (generally referred as {\it
    superequilibration}~\cite{Chung:2009qs}) is ensured when reactions
  like $\tilde \ell\tilde \ell \to \ell\ell$ are faster than the
  Universe expansion rate.  These reactions are induced by gaugino
  interactions, but since they require a gaugino chirality flip they
  turn out to be proportional to its mass $m_{\tilde g}$, and 
  can be neglected in the limit  $m_{\tilde g}\to 0$.

  Furthermore, since the $\mu$ parameter of the $H_u H_d$
  superpotential term is expected to be of the order of the soft
  gaugino masses, it is reasonable to consider in the same temperature
  range also the effect of the higgsino mixing term, which implies
  that the sum of the up- and down- higgsino chemical potentials
  vanishes.  The rates of the corresponding reactions, given
  approximately by $\Gamma_{\tilde g} \sim m^2_{\tilde g}/T$ and
  $\Gamma_{\mu} \sim \mu^2/T$, are faster than the Universe expansion
  rate up to temperatures
\be
\label{eq:Tgmu}
 T\lsim 5\cdot 10^7 
\left(\frac{m_{\tilde g},\,\mu}{500\,{\rm GeV}}\right)^{2/3}\, {\rm GeV}.
\ee
The corresponding chemical equilibrium relations enforce the
conditions:
\bea
\label{eq:geq0}
\tilde g &=&0, \\ 
\label{eq:mueq0}
\tilde H_u+\tilde H_d&=&0.  
\eea
\item[$7_{\rm SE}$.]  Up to temperatures given by~\eqref{eq:Tgmu} the
  MSSM has the same global anomalies than the SM, that are the EW
  $SU(2)$-$U(1)_{B+L}$ mixed anomaly and the QCD chiral anomaly. They
  generate the effective operators $O_{EW}=\Pi_\alpha(QQQ\ell_\alpha)$
  and $O_{QCD}=\Pi_i(QQu^c_{Li}d^c_{Li})$. Above the EW phase
  transition reactions induced by these operators are in thermal
  equilibrium, and  the corresponding conditions read:
\bea
\label{eq:EW}
&&9\,Q+\sum_\alpha \ell_\alpha = 0 \\
\label{eq:QCD}
&&6\,Q-\sum_i\left(u_i+d_i\right)=0\,,
\eea
where we have used the same chemical potential for the three quark doublets
(Eq.\eqref{eq:Q}), which is always appropriate in the SE regime below the limit
\eqref{eq:Tgmu}.

\end{enumerate}

\medskip
 \subsubsection{Flavour charges}
\label{sec:DeltaAlpha}

Eqs.\eqref{eq:leptons} and \eqref{eq:upquarks}--\eqref{eq:Q}, together
with the SE conditions \eqref{eq:geq0}-\eqref{eq:mueq0}, the two
anomaly conditions \eqref{eq:EW}-\eqref{eq:QCD} and the hypercharge
neutrality condition \eqref{eq:Ytotmu}, give $11+2+2+1=16$ constraints
for the 18 chemical potentials.  Note however that there is one
redundant constraint, that we take to be the QCD sphaleron condition,
since by summing up Eqs.\eqref{eq:upquarks} and \eqref{eq:downquarks}
and taking into account \eqref{eq:Q}, \eqref{eq:geq0}, and
\eqref{eq:mueq0} we obtain precisely Eq.\eqref{eq:QCD}. Therefore,
like in the SM, we have three independent chemical potentials, that
could be taken to be the ones corresponding to the leptons
doublets. Another choice, that is more useful in leptogenesis, is to
define three linear combinations of the chemical potentials
corresponding to the $SU(2)$ anomaly free flavour charges
$\Delta_\alpha\equiv B/3-L_\alpha$. The reason is that these charges,
being anomaly free and perturbatively conserved by the low energy MSSM
Lagrangian, evolve {\it slowly} because the corresponding symmetries
are violated only by the heavy Majorana neutrino dynamics. Their
evolution is thus determined by reactions belonging to class III, and
needs to be computed by means of three independent Boltzmann
equations.  We define the number densities of particles per degree of
freedom, normalized to the entropy density $s$, as $Y_{\Delta b,\Delta
  f} \equiv \frac{1}{g_{b,f}} \frac{\Delta n_{b,f}}{s}$.  In terms of
these quantities the density of the $\Delta_\alpha$ charges normalized
to the entropy density can be written as:
\be 
\label{eq:YDeltaAlpha}
Y_{\Delta_\alpha} = 3\,\left[\frac{1}{3}
    \sum_i\left(2Y_{\Delta Q_i}+Y_{\Delta u_i}+Y_{\Delta d_i}\right)-
  (2Y_{\Delta \ell_\alpha}+Y_{\Delta
    e_\alpha})-\frac{2}{3}Y_{\Delta\tilde g}\right]\,. 
\ee
The expression above is completely general and holds in all
temperature regimes, including the NSE regime (see Section
\ref{sec:NSE}).  Note that $\tilde g$ in the equation above cancels
for the quarks but not for the leptons, and thus in the NSE, in which
the gaugino chemical potential does not vanish, when the
$Y_{\Delta_\alpha}$ charges are expressed just in terms of the 
number density asymmetries of the fermions, $Y_{\Delta\tilde g}$ 
also contributes.

In Eq.\eqref{eq:YDeltaAlpha} we have left in clear some numerical
factors: the overall factor of 3 adds the contributions of scalars
(that is twice that of fermions), the factor of 2 in front of the
$Y_{\Delta Q_i}$ and $Y_{\Delta \ell_\alpha}$ accounts for the $SU(2)$
gauge multiplicity, while the color factor compensates against the the
quark baryon number $B=1/3$.

The density asymmetries of the doublet leptons and higgsinos, that
weight the washout terms in the Boltzmann equations, can now be
expressed in terms of the anomaly free charges by means of the $A$
matrix and $C$ vectors introduced respectively in ref.~\cite{barbieri}
and ref.~\cite{flavour2} that are defined as:
\be
\label{eq:AC}
Y_{\Delta\ell_\alpha}= A^\ell_{\alpha\beta}\,\, Y_{\Delta_\beta}, 
\qquad\qquad 
Y_{\Delta \tilde H_{u,d}}= 
C^{\tilde H_{u,d}}_\alpha\,\,  Y_{\Delta_\alpha}. 
\ee
Here and in the following we will give results for the $A$ and $C$
matrices for the fermion states.  We recall that in the SE regime the
density asymmetry of a scalar boson that is in chemical equilibrium with
its fermionic partner is given simply by $Y_{\Delta b}=2\,Y_{\Delta f}$
with the factor of 2 from statistics.

\subsubsection{All Yukawa reactions in equilibrium}
\label{sec:ACT1}

Assuming moderate values of $\tan\beta$, at temperatures below the
limit in Eq.\eqref{eq:Te} standard leptogenesis cannot be
successful. However, this range can correspond to a low temperature
windows in which soft leptogenesis~\cite{soft1,soft2,soft3} can successfully
proceed, and therefore it is worth giving the results for $A$ and
$C$. They are: 
\begin{eqnarray}
A^\ell=\frac{1}{9\times 237}\left(
\begin{array}{ccc}
-221 & 16  & 16\\
 16 & -221 &  16 \\
 16 &  16  & -221
\end{array}\right), 
&\;\;\;\;\;& 
C^{\tilde H_u}=-C^{\tilde H_d}=\frac{-4}{237}\left(1,\;1,\;1\right).
\label{eq:ACT1}
\end{eqnarray}
Note that since in this regime the chemical potentials for the scalars
and leptons degrees of freedom of each chiral multiplet equilibrate,
the analogous results for
$Y_{\Delta\ell_\alpha}+Y_{\Delta\tilde\ell_\alpha}$ can be obtained by
simply multiplying the $A$ matrix in Eq.\eqref{eq:ACT1} by a factor of
3. This gives the same $A$ matrix obtained in the non-supersymmetric
case in the same regime (see e.g. eq.(4.13) in ref.~\cite{flavour2}).
The $C$ matrix (multiplied by the same factor of 3) differs from the
non-supersymmetric result by a factor $1/2$. This is because after
substituting $\tilde H_d=-\tilde H_u$ (see Eq.\eqref{eq:mueq0}) all
the chemical potential conditions are formally the same than in the SM
with $\tilde H_u$ identified with the chemical potential of the scalar
Higgs, but since $C$ expresses the result for number densities, in the
SM a factor of 2 from boson statistics appears for the SM Higgs.  This
agrees with the analysis in ref.\cite{inui}, and is a general result
that holds for supersymmetry within the SE regime.

\subsubsection{Electron and up-quark Yukawa reactions out of
  equilibrium}
\label{sec:ACT2}

Raising the temperature above $10^5(1+\tan^2\beta)\,{\rm GeV}$ the
interactions mediated by the electron Yukawa $h_e$ are not able to
maintain equilibrium, and one condition in Eq.\eqref{eq:leptons} for
$\alpha=e$ is lost.  However, since in the effective theory at this
temperature one can set $h_e\to 0$, one global symmetry is
gained. This corresponds in the fermion sector to chiral symmetry for
the R-handed electron, that in the present case translates into a
symmetry under phase rotations of the $e$ chiral multiplet that holds
in the limit of unbroken supersymmetry.  Conservation of the
corresponding charge ensures that $\Delta n_{e}+\Delta n_{\tilde e} =3
\Delta n_{e}$ is constant, and since leptogenesis aims to explain
dynamically the generation of a lepton asymmetries we set this
constant to zero, so that the R-handed electron chemical potential is
$e=0$.  In this way the chemical equilibrium condition that is lost is
replaced by a new condition implied by the conservation of a global
charge, and three independent chemical potentials (or alternatively
the three non-anomalous charges \eqref{eq:YDeltaAlpha}) are still
sufficient to describe all the density asymmetries of the
thermodynamic system.  At temperatures above $T\sim 2\cdot 10^6\,$GeV
interactions mediated by the up-quark Yukawa coupling $h_u$ drop out
of equilibrium.  In this case however, by setting $h_u \to 0$ no new
symmetry is obtained, since chiral symmetry for the R-handed quarks is
anomalous and the corresponding charge is not conserved by fast QCD
sphaleron interactions. However, after dropping the first condition in
Eq.\eqref{eq:upquarks} for $u_i=u$, the QCD sphaleron condition
Eq.\eqref{eq:QCD} ceases to be a redundant constraint, with the result
that also in this case no new chemical potentials are needed to
determine all the particle density asymmetries.  The corresponding
results, that can be again relevant for soft leptogenesis, read:
\begin{eqnarray}
A^\ell=\frac{1}{3\times 2886}\left(\begin{array}{ccc}
-1221 &\ \,\, 156  &\ \,\, 156 \\
\ \, 111 &-910  &\ \, 52\\
\ \, 111 &\ \, 52  &-910
\end{array}\right), &\;\;\;& C^{\tilde H_u}=-C^{\tilde H_d}=
\frac{-1}{2886}\left(37,\;52,\;52\right).
\label{eq:ACT2}
\end{eqnarray}

\subsubsection{First generation Yukawa reactions out of equilibrium
  (SE regime)}
\label{sec:ACT3SE}

Let us now consider what happens at temperatures $T\gsim 4\cdot
10^6(1+\tan^2\beta)\,{\rm GeV,} $ when also the $d$-quark Yukawa coupling
can be set to zero (in order to remain within the SE regime we assume
$\tan\beta \sim 1$).  In this case the equilibrium dynamics is
symmetric under the exchange $u \leftrightarrow d$ (both chemical
potentials enter only the QCD sphaleron condition Eq.\eqref{eq:QCD} with
equal weights) and so must be any physical solution of the set of
constraints. Thus, the first condition in Eq.\eqref{eq:downquarks} can be
replaced by the condition $d=u$, and again three independent
quantities suffice to determine all the particle density asymmetries.
The corresponding result is:
\begin{eqnarray}
A^\ell=\frac{1}{3\times 2148}
\left(\begin{array}{ccc}
-906 &\ \,\, 120  &\ \,\, 120\\
\ \, 75 &-688 &\ \, 28\\
\ \, 75 &\ \,28  &-688 
\end{array}\right), &\;& C^{\tilde H_u}=-C^{\tilde H_d}=
\frac{-1}{2148}\left(37,\;52,\;52\right), \qquad
\label{eq:ACT3SE}
\end{eqnarray}
that agree with what is obtained in non-supersymmetric leptogenesis
(see eq.~(4.12) of ref.~\cite{flavour2}) after the factor 1/2 for the
higgsinos discussed below \Eqn{eq:ACT1} is accounted for.  In the
numerical analysis in Section~\ref{sec:results} we will take this case
as a benchmark to confront the results obtained in the SE regime with
the corresponding results in the NSE regime.

\medskip

\subsection{Non-superequilibration Regime}
\label{sec:NSE}

At temperatures above the limit given in Eq.\eqref{eq:Tgmu} the Universe
expansion is fast enough that reactions induced by $m_{\tilde g}$ and
$\mu$ do not occur. Setting to zero in the high temperature effective
theory these two parameters has the following consequences:
\begin{enumerate}
\item[(i)] Condition \eqref{eq:geq0} has to be dropped, and gauginos
  acquire a non-vanishing chemical potential $\tilde g\neq 0$
  (corresponding to the difference between the number of L and R
  helicity states). The chemical potentials of the members of the same
  matter supermultiplets are no more equal (non-superequilibration)
  but related as in Eqs.\eqref{eq:tQtell}--\eqref{eq:tutdte}.

\item[(ii)] Condition \eqref{eq:mueq0} also has to be dropped, and the
  chemical potentials of the up- and down-type Higgs and higgsinos do
  not necessarily sum up to zero.

\item[(iii)] The MSSM gains two new global symmetries: $m_{\tilde
    g}\to 0$ yields a global $R$-symmetry, while $\mu\to 0$
  corresponds to a  global symmetry of the Peccei-Quinn (PQ)
  type.\footnote{We assume that, similarly to $m_{\tilde g}$, in this
    regime also the other soft supersymmetry breaking terms can be neglected,
    and in particular the $B$-term for the heavy sneutrinos. This is
    always the case in the high temperature regime for successful
    leptogenesis.  However, at the lower temperatures required in soft
    leptogenesis this is not necessarily true.  Then the $R$-charge is
    not perturbatively conserved, as is clear from the last entry in
    Table~\ref{tab:2}. This can have quite interesting consequences
    for soft leptogenesis that are analyzed in a companion
    paper~\cite{inpreparation}.}
\end{enumerate}

\subsubsection{Anomalous and non-anomalous symmetries}
\label{sec:symmetries}

The charges of the various states under the $R$ and $PQ$ symmetries
together with the values of the other two global symmetries $B$ and
$L$ are given in Table~\ref{tab:1}.  Like $L$, also $R$ and $PQ$ are
not symmetries of the seesaw superpotential terms $MN^cN^c+ \lambda
N^c\ell H_u$.  In Table~\ref{tab:1} we have fixed the charges of the
heavy $N^c$ supermultiplets in such a way that the mass term $MN^cN^c$
is invariant under all symmetries and in particular has $R$ charge
equal to 2.\footnote{Under $R$-symmetry the superspace Grassmann
  parameter transform as $\theta \to e^{i\alpha}\theta$ . Invariance
  of $\int d\theta\, \theta =1$ then requires $R(d\theta)=-1$. Then
  the chiral superspace integral of the superpotential $\int
  d\theta^2\, W $ is invariant if $R(W)=2$. By expanding a chiral
  supermultiplet in powers of $\theta$ it is also clear that its $R$
  charge equals the charge of the scalar boson $R(b)=R(f)+1$.}  In
contrast, the Yukawa term $N^cLH_u$ violates $L$, $PQ$ and $R$.  All
the four global symmetries $B$, $L$, $PQ$ and $R$ have mixed gauge
anomalies with $SU(2)$, and $R$ and $PQ$ have also mixed gauge
anomalies with $SU(3)$.  Two linear combinations $R_2$ and $R_3$ of
$R$ and $PQ$, having respectively only $SU(2)$ and $SU(3)$ mixed
anomalies have been identified in ref.~\cite{Ibanez:1992aj}. They
are:~\footnote{For definiteness we restrict ourselves to the case of
  three generations $N_g=3$ and one pair of Higgs doublets $N_h=1$.
  We also normalize $R_{2,3}$ in such a way that $R_{2,3}(b)
  =R_{2,3}(f)+1$.}
\bea
\label{eq:R2}
R_2 &=& R-2\,PQ \\
\label{eq:R3}
R_3 &=&R-3\,PQ\,. 
\eea
The values of $R_{2,3}$ for the different states are given in
Table~\ref{tab:2}.  The authors of ref.~\cite{Ibanez:1992aj} have also
constructed the effective multi-fermions operators generated by the
mixed anomalies:
\bea
\label{eq:tO-EW}
\tilde O_{EW} &=&
\Pi_\alpha \left(QQQ\ell_\alpha\right)\; \tilde H_u\tilde H_d\;\tilde W^4\,,\\
\label{eq:tO-QCD}
\tilde O_{QCD} &=& \Pi_i \left(QQu^c d^c\right)_{i}\; \tilde g^6 \,.
\eea
 \begin{table}[t!!]
 \begin{center} 
% \begin{tabular}{|c|c|c|c|c|c|c|c|c|c|}
\begin{tabular}{|cc|c|c|c|c|c|c|c|c|c|c|c|c|}
 \hline
&
&\quad  $\tilde g\ $
&\quad $Q\ $
&\quad$u^c \ $
&\quad $d^c \ $
&\quad $\ell \ $
&\quad $e^c \ $
&\quad $\tilde H_d\phantom{\Big|}$
&\quad$\tilde H_u\ $
&\quad$N^c\  $
\\
\hline
\multicolumn{2}{|c|}{$B\phantom{\Big|}$} &$ 0$&$ \frac{1}{3}$&$-\frac{1}{3}$&$-\frac{1}{3}$&$ 0$&$ 0$&$ 0$&$ 0$& 0 \\ \hline 
\multicolumn{2}{|c|}{$L\phantom{\Big|}$} &$ 0$&$ 0$          &$ 0$          &$ 0$          &$ 1$&$-1$&$ 0$&$ 0$& 0 \\ \hline 
\multicolumn{2}{|c|}{$PQ\phantom{\Big|}$}&$ 0$&$ 0$          &$-2$          &$ 1$          &$-1$&$ 2$&$-1$&$ 2$& 0 \\ \hline 
\multirow{2}{*}{$R$}&$\phantom{\!\!\!\!\Big|}f$                 &$ 1$&$-1$          &$-3$          &$ 1$          &$-1$&$ 1$&$-1$&$ 3$& 0 \\ \cline{3-11}
 &$\phantom{\!\!\!\!\Big|}b$             &$ 2$&$ 0$          &$-2$          &$ 2$          &$ 0$&$ 2$&$ 0$&$ 4$& $1$ \\ \hline
\end{tabular}
\caption{$B$, $L$, $PQ$ and $R$ charges for the particle supermultiplets 
  that are labeled in the top row by their L-handed fermion
  component.  Note that we use chemical potentials for the R-handed
  $SU(2)$ singlet fields $u,\,d,\,e$ that have opposite charges with
  respect to the ones for $u^c,\,d^c,\,e^c$ given in the
  table. The $R$-charges for bosons are determined by $R(b)=R(f)+1$.}
\label{tab:1}
\end{center}
\end{table}

Given that three global symmetries $B$, $L$ and $R_2$ have mixed
$SU(2)$ anomalies (but are free of $SU(3)$ anomalies) it is clear that
we can construct 2 anomaly free combinations, the first one of which
we chose to be, as in the SM, $B-L$. Given that only one term, that is
$N^c\,\ell H_u$, violates perturbatively the global symmetries, a
second anomaly free combination ${\cal R}$ can be chosen in such a way
that it is respected also by the corresponding interactions, and thus
it is an exact symmetry of the MSSM+seesaw in the NSE regime.  The
corresponding charge reads
\be
\label{eq:calR}
 {\cal R}=\frac{5}{3}B-L+R_2,  
\ee
and is exactly conserved.  In the $SU(3)$ sector, besides the chiral
anomaly we now have also $R_3$ mixed anomalies. Thus also in this
case anomaly free combinations can be constructed, and  in particular we
can define one combination for each quark superfield.
Assigning to the L-handed supermultiplets chiral
charge $\chi=-1$ these combinations have the form:
% is $\chi+6\,R_3$, but other (more useful) combinations can be constructed
% which involve chiral rotations of a single fermion multiplet. 
%
\be
\label{eq:chiral} 
\chi_{q_L}+\kappa_{q_L}\,R_3
\ee
where, for example, $\kappa_{u^c_L}=\kappa_{d^c_L}=1/3$ and
$\kappa_{Q_L}=2/3$.  Note that since $R_3$ is perturbatively conserved
by the complete MSSM+seesaw Lagrangian, when the Yukawa coupling of
one quark is set to zero the corresponding charge \Eqn{eq:chiral}
will be exactly conserved.

\subsubsection{Constraints in the non-superequilibration regime}
\label{sec:NSEconstraints}

As explained in the introduction to this section, conditions
corresponding to conservation laws constrain particle number
density-asymmetries, while conditions from reactions that are in
chemical equilibrium constrain particle chemical potentials.
Eventually, to get a closed form solution to the set of constraining
equations one needs to use a single set of variables. In the
following, we will then express the number density-asymmetries of
relativistic particles in terms of their chemical potentials by means
of \eqn{eq:Dnmu} that holds in the ultra-relativistic limit and at
first order in $\mu_{f,b}/T$.
%\footnote{Non relativistic corrections and/or higher order
%  terms in $\mu_{f,b}$ could be straightforwardly included; they are
%  however, numerically irrelevant and would prevent obtaining simple
%  algebraic solutions to the set of constraining equations.}

In the NSE regime, the conditions listed in items
$6_{SE}$ and $7_{SE}$ of the previous section have to be dropped, but
new conditions arise.

\begin{enumerate}

\item[$6_{NSE}$.]  The conservation law for the 
${\cal R}$ charge yields the following global neutrality condition:
 \begin{eqnarray}
   \nonumber
 {\cal R}_{\rm tot} &=&  
\sum_f \Delta n_f  {\cal R}_f + 
\sum_b \Delta n_b  {\cal R}_b +
\Delta n_{\tilde N_1}  {\cal R}_{\tilde N_1}\\
 &=& \frac{T^2}{6}\left(
\sum_{f} \mu_f\,g_f\,{\cal R}_f + 
2\,\sum_{b} \mu_b\,g_b\, {\cal R}_b \right) 
-\Delta n_{\tilde N_1}
=0,\qquad \quad
    \label{eq:calRtot}
  \end{eqnarray}
with the term in parenthesis given by:
\bea
\nonumber
&&\!\!\!\sum_{f} \mu_f\,g_f\,{\cal R}_f + 
2\,\sum_{b} \mu_b\,g_b\, {\cal R}_b = \\
&& \qquad 
2\,\left(\sum_i\left(2Q_i-5u_i+4d_i\right)+ 
2\sum_\alpha\left(\ell_\alpha+e_\alpha\right)+5\tilde H_d-\tilde H_u
+31\,\tilde g\right).  
\label{eq:calRtotmu}
\eea
The last terms in both lines of \Eqn{eq:calRtot} correspond to the
contribution to ${\cal R}$-neutrality from the lightest sneutrino
asymmetry $\Delta n_{\tilde N_1} = n_{\tilde N_1}- n_{\tilde N_1^*}$
with charge ${\cal R}_{\tilde N_1}= - {\cal R}_{N^c}=-1$.  All the
other heavier neutrinos are assumed to have already decayed at the
time $N_1$ leptogenesis takes place, and do not contribute.  Note that
since in general $\tilde N_1$ is not in chemical equilibrium, no
chemical potential can be associated to it, and hence this constraint
needs to be formulated in terms of its number density asymmetry that
has to be evaluated by solving a Boltzmann equation for
$Y_{\Delta_{\tilde N}} \equiv Y_{\tilde N_1}- Y_{\tilde N_1^*}$ (see
Section 3).  Note also that since the final lepton asymmetry in
leptogenesis is generally determined by the dynamics at $T<M$, where
the sneutrino abundance gets exponentially suppressed, we can expect
that neglecting $\Delta n_{\tilde N_1}$ in \Eqn{eq:calRtot} could
still give a good approximation. This is indeed confirmed by the
numerical analysis presented in Section~\ref{sec:results}.

\item[$7_{NSE}$.] The operators in Eqs.\eqref{eq:tO-EW}--\eqref{eq:tO-QCD} induce
  transitions that in the NSE regime are in chemical equilibrium. 
This enforces the generalized EW and QCD sphaleron equilibrium
conditions~\cite{Ibanez:1992aj}: 
\bea
\label{eq:tEWmu}
&&3\sum_i Q_i+\sum_\alpha\ell_\alpha
+\tilde H_u+\tilde H_d+4\,\tilde g=0,\\
\label{eq:tQCDmu}
&&2\sum_i Q_i-\sum_i\left(u_i+d_i\right)+6\,\tilde g=0, 
\eea
that replace \eqref{eq:EW} and \eqref{eq:QCD}.

\item[$8_{NSE}$.] The chiral-$R_3$ charges in Eq.\eqref{eq:chiral} are
  anomaly free, but clearly they are not conserved by the quarks
  Yukawa interactions. However, when a quark supermultiplet decouples
  from its Yukawa interactions an exact conservation law arises. (Note
  that $h_{u,d} \to 0$ implies $u$ and $d$ decoupling, but $Q_1$
  decoupling is ensured only if also $h_{c,s}\to 0$.)  
  The conservation laws corresponding to these symmetries read:
\bea
\label{eq:chiralmuquarks}
\frac{T^2}{6}\left[3 q_R + 6(q_R-\tilde g)\right]+\frac{1}{3}
\,R_{3\;\rm tot}
&=&0 \\
\label{eq:chiralmuquarks2}
\frac{T^2}{6}\,2\,\left[3 Q_L + 6(Q_L+\tilde g)\right]-\frac{2}{3}
\,R_{3\;\rm tot}
&=&0
\eea  
and hold for $q_R=u_i,\,d_i$ and $Q_L=Q_i$ in the regimes when the
appropriate Yukawa reactions are negligible.  Note the factor of 2 for
the $Q_L$ chiral charge in front of the first square bracket in
\Eqn{eq:chiralmuquarks2} that is due to $SU(2)$ gauge multiplicity.
In terms of chemical potentials and of the sneutrino number density
asymmetry, the total $R_3$ charge in \Eqns{eq:chiralmuquarks}{eq:chiralmuquarks2} reads:
\be
\label{eq:R3tot}
R_{3\;\rm tot}=\frac{T^2}{6\,}
\left(\sum_{f} \mu_f\,g_f\,{R_3}_f + 
2 \sum_{b} \mu_b\,g_b\, {R_3}_b\right) +\Delta n_{\tilde N_1}  
{R_3}_{\tilde N_1} 
\ee
where ${R_3}_{\tilde N_1} =-1$, and the 
quantity in parenthesis is given by:
\bea
\nonumber 
&&\!\!\!\!\!\!\!
\sum_{f} \mu_f\,g_f\,{R_3}_f + 
2 \sum_{b} \mu_b\,g_b\, {R_3}_b = \\
&& 82\,\tilde g -3\sum_i \left(2\,Q_i+11\,u_i-4\,d_i\right)
+ 
\sum_\alpha\left(16\,\ell_\alpha+13\,e_\alpha\right)
+16\,\tilde H_d-14\,\tilde H_u. \quad
\label{eq:R3totmu}
\eea

As regards the leptons, since they do not couple to the QCD anomaly,
by setting $h_{e}\to 0$ a symmetry under chiral supermultiplet
rotations is directly gained for the R-handed leptons implying $\Delta
n_e+\Delta {\tilde n_e}=0$ and giving the condition:
\be
\label{eq:chiralmuleptons}
e-\frac{2}{3}\tilde g =0.
\ee
No analogous condition arises for the lepton doublets relevant for
leptogenesis, since by assumption they remain coupled via Yukawa
couplings to the heavy $N$'s.
\end{enumerate}

Let us now evaluate what is the number of independent chemical
potentials in the NSE regime. Since we already know that this number
should not change when some constraints from chemical equilibrium are
dropped and replaced by conservation conditions, let us start by
assuming $\tan\beta$ large enough that the Yukawa couplings $h_e$ and
$h_d$ enforce the equilibrium conditions \eqref{eq:leptons} and
\eqref{eq:downquarks}, and also keep for the time being the condition
\eqref{eq:upquarks} for the up-quark.  We have 9 conditions from the
Yukawa couplings plus 2 from quark doublets equilibration
\eqref{eq:Q}, the generalized EW sphaleron condition \eqref{eq:tEWmu}
and the QCD sphaleron condition \eqref{eq:tQCDmu} that now must be
counted since it is independent from other constraints, the global
hypercharge \eqref{eq:Ytotmu} and $\cal R$ \eqref{eq:calRtotmu}
neutrality conditions, for a total of 15 constraints for the 18
chemical potentials or corresponding particle density asymmetries
considered in the previous section.  However, in the NSE regime we
have that the set of constraining conditions involve one additional
independent quantity that is the sneutrinos density asymmetry
$Y_{\Delta_{\tilde N}}$, and therefore the total number of relevant
quantities is 19. We thus conclude that in the NSE regime the three
flavour charges $Y_{\Delta_\alpha}$ in Eq.\eqref{eq:YDeltaAlpha} do
not suffice to describe the density asymmetries of all MSSM particles
and superparticles, and that $Y_{\Delta_{\tilde N}}$ is also required.

To proceed, we can now set $h_u\to 0$ to match the actual number of
equilibrium constrains in the NSE regime. We lose one Yukawa
condition, but we gain the corresponding chiral-$R_3$ conservation
condition \eqref{eq:chiralmuquarks} and, as expected, the overall
number of independent chemical potentials or number density
asymmetries remains the same.

 \begin{table}[t!]
\begin{center}
\begin{tabular}{|cc|c|c|c|c|c|c|c|c|c|c|c|c|}
 \hline
&
&\quad  $\tilde g\ $
&\quad $Q\ $
&\quad$u^c \ $
&\quad $d^c \ $
&\quad $\ell \ $
&\quad $e^c \ $
&\quad $\tilde H_d\phantom{\Big|}$
&\quad$\tilde H_u\ $
&\quad$N^c\  $
\\
\hline
\multirow{2}{*}{$R_2$} &$\phantom{\!\!\!\!\Big|}f$     &$ 1$&$-1$&$ 1$&$-1$&$ 1$&$ -3$&$1$&$-1$& 0\\ \cline{3-11}
                     &$\phantom{\!\!\!\!\Big|}b$       &$ 2$&$ 0$&$ 2$&$ 0$&$ 2$&$-2$&$ 2$&$ 0$& 1 \\ \hline
\multirow{2}{*}{$R_3$} &$\phantom{\!\!\!\!\Big|}f$     &$ 1$&$-1$&$ 3$&$-2$&$ 2$&$ -5$&$2$&$-3$& 0\\ \cline{3-11}
                     &$\phantom{\!\!\!\!\Big|}b$       &$ 2$&$ 0$&$ 4$&$ -1$&$ 3$&$-4$&$ 3$&$-2$& 1 \\ \hline
\hline
\multirow{2}{*}{$\cal R$}&$\phantom{\!\!\!\!\Big|}f$ 
&$ 1$&$-\frac{4}{9}$&$\frac{4}{9}$&$-\frac{14}{9}$&$0$&$ -2$&$ 1$&$-1$& 0 \\ \cline{3-11}
                     &$\phantom{\!\!\!\!\Big|}b$       
&$ 2$&$\frac{5}{9}$&$\frac{13}{9}$&$-\frac{5}{9}$&$ 1$&$-1$&$ 2$&$ 0$& 1 \\ 
\hline
\end{tabular}
\caption{Charges for the fermionic and bosonic components of the SUSY
  multiplets under the $R$-symmetries defined in Eqs.\protect
  \eqref{eq:R2}, \protect \eqref{eq:R3} and \protect \eqref{eq:calR}.
  Supermultiplets are labeled in the top row by their L-handed fermion
  component.  We use chemical potentials for the R-handed $SU(2)$
  singlet fields $u,\,d,\,e$ that have opposite charges with respect
  to the ones for $u^c,\,d^c,\,e^c$ given in the table.}
\label{tab:2}
\end{center}
\end{table}

\medskip

In the following we give some results corresponding to temperature
ranges relevant for successful supersymmetric leptogenesis, that
always occurs within the NSE regime.  Given that the Yukawa conditions for
the leptons and for the down-type quarks depend on the value of
$\tan\beta$, the temperature ranges we refer to is only
indicative (and correspond to moderate values of $\tan\beta$), but we
will always specify in a clear way which conditions are used to obtain
the corresponding results.  In the non-superequilibration regime there
are different flavour mixing matrices for the scalar and fermion
components of the leptons and Higgs supermultiplets.  To express more
concisely all the results, it is convenient to introduce a new $C$
vector to describe the gaugino number density asymmetry 
per degree of freedom in terms of the relevant charges:
\bea
\label{eq:Cg}
Y_{\Delta \tilde g}=C_a^{\tilde g}\;Y_{\Delta_a}\, ,  
&\;\;\;\;\;\;\;{\rm with}\;\;\;\;\;\;\;\;\;
&\Delta_{a}=\left(\Delta_\alpha,\Delta_{\tilde N}\right)\; .
\eea
% 
%According to \eqref{eq:tQtell} and \eqref{eq:HuHd} and taking into account
%the statistical factor of 2 for bosons with respect to fermions, the
%corresponding matrices for the scalars are then given by $A^{\tilde
%  \ell}_{\alpha\beta} = 2\,(A^{\ell}_{\alpha\beta}+C^{\tilde
%  g}_\beta)$ and $C^{H_{u,d}}= 2\,(C^{\tilde H_{u,d}}+C^{\tilde g})$.

\subsubsection{First generation Yukawa reactions out of equilibrium
  (NSE regime).} 
\label{sec:ACT3NSE}

In the temperature range between $10^8$ and $10^{11}\,$GeV, and for
moderate values of $\tan\beta$, all the first generation Yukawa
couplings can be set to zero.  Using for $u,\,d$ 
conditions \eqref{eq:chiralmuquarks} and for $e$ 
condition \eqref{eq:chiralmuleptons} as are implied  
by $h_{u,d},\,h_e\to 0$ we obtain:  
\begin{eqnarray}
\nonumber
A^\ell&=&\frac{1}{9\times 162332}\left(
\begin{array}{cccc}
-198117 &\ \,\, 33987  &\ \,\, 33987& -8253 \\
\ \, 26634 &-147571  &\ \, 14761&-8055\\
\ \, 26634 &\ \, 14761   &-147571&-8055
\end{array}
\right), \\ [5pt]
\nonumber
C^{\tilde g}&=&\frac{-11}{162332}\left(163,\;
165,\;165,\;-255\right),\quad \\ [5pt]
\nonumber
C^{\tilde H_u}&=& 
\frac{-1}{162332}\left(3918,\;
4713,\; 4713,\; 95 \right)\,, \\ [5pt]
C^{\tilde H_d} &=& \frac{1}{3\times 162332  }\left(5413,\;
9712,\;9712,\;-252\right)\,,  
\label{eq:ACT3NSE}
\end{eqnarray}
where the rows correspond to $( Y_{\Delta_e},\; Y_{\Delta_\mu},\;
Y_{\Delta_\tau},\; Y_{\Delta_{\tilde N}})$.  For completeness, in
\Eqn{eq:ACT3NSE} we have also given the results for $C^{\tilde H_d}$
even if only the up-type Higgs density asymmetry is relevant for the
leptogenesis processes.  Note that neglecting the contribution of
$\Delta n_{\tilde N_1}$ to the global charges ${\cal R}_{\rm tot}$ in
\Eqn{eq:calRtot} and ${R_3}_{\rm tot}$ in \Eqn{eq:R3tot} corresponds
precisely to set to zero the fourth column in all the previous
matrices. Then, analogously with the SE and SM cases, within this
`3-columns approximation' all particle density asymmetries can be
expressed just in terms of the three $Y_{\Delta_\alpha}$ charge
densities.

In the numerical analysis carried out in the next section we will
confront the results obtained with the full matrices
\eqref{eq:ACT3NSE} with the results obtained for the analogous case
discussed in Section \ref{sec:ACT3SE} of first generation Yukawas out
of equilibrium, but within the SE regime, that yielded the matrices in
\eqref{eq:ACT3SE}.

\subsubsection{Second generation Yukawa reactions 
out of equilibrium: two flavour regime.} 
\label{sec:ACT4}

In the temperature range between $10^{11}$ and $10^{12}\,$GeV we set
$h_\mu\to 0$. This condition defines the two flavour regime for
leptogenesis~\cite{barbieri,flavour1,flavour2}.  In this regime we
have also $h_s,h_c\to 0$ that result in conserved
$\chi_{q^c_L}+\kappa_{q^c_L} R_3$ charges ($q=s,\,c$) yielding the
corresponding conditions \eqref{eq:chiralmuquarks}.  Since the first
generation quark doublet does not mix with the other generations,
$Q_1$ is not equal to $Q=Q_2=Q_3$, but conservation of
$\chi_{Q_1}+\kappa_{Q_1}R_3$ then provides the required new condition
\eqref{eq:chiralmuquarks2}.  Conservation laws stemming from
$h_{e,\mu,u,d,s,c}\to 0$ then add up to a total of 7 constraints that
replace the 6 Yukawa equilibrium conditions plus the condition
$Q_1=Q$.

In the two flavour regime only the lepton doublet $\ell_\tau$ is
identified, while the $\ell_e$ and $\ell_\mu$ doublets, both of which
have no Yukawa interactions, are not distinguished. In general, one
combination $\ell_{e\mu}$ of these two doublets remains coupled to
lightest heavy singlet $N_1$, while the orthogonal combination
decouples, so that $N_1$ decays do not generate directly a lepton
asymmetry in this particular flavour direction.  The corresponding
non-anomalous $B/3-L_\alpha$ charge is thus exactly conserved, and we
can set its value to zero. This condition, together with the third
generation Yukawa conditions, the two sphaleron conditions, and $\cal
Y$ and $\cal R$ global neutrality, give 8 constraints for the
remaining 10 chemical potentials
$Q,\,b,\,t,\,\ell_\alpha,\tau,\,\tilde H_{u,d},\,\tilde g$.
Therefore, in this case the two flavour charge densities
$Y_{\Delta_{e\mu}}$ and $Y_{\Delta_\tau}$ together with $Y_{\Delta_{\tilde N}}$ suffice to express the
density asymmetries for all the other particles. In the basis
$(Y_{\Delta_{e\mu}},\; Y_{\Delta_\tau},\;Y_{\Delta_{\tilde N}})$ we obtain:
%{\bf agreed by all after corrections in v5}:
%
\begin{eqnarray}
\nonumber
A^\ell&=&\frac{1}{6\times 580163}\left(
\begin{array}{ccc}
-460047 &\ \,\, 88166&-17229\\
\ \, 80783 &-349740&-14094  \\
\end{array}
\right),\\ [5pt]
\nonumber
C^{\tilde g}&=&\frac{1}{2\times 580163}\left(-11056,\;
-11786,\;20307\right), \\ [5pt]
\nonumber
C^{\tilde H_u}&=&
\frac{-1}{2\times 580163}
\left(69131,\; 70652,\;4579\right), \\ [5pt]
C^{\tilde H_d} &=& \frac{1}{6\times 580163}
\left(8939,\;77012,\;-12483\right). 
\label{eq:ACT4}
\end{eqnarray}
Note that since EW sphaleron flavour mixing involves all the three
lepton doublets, a non-vanishing lepton asymmetry (equal to $B/3$) is
induced also in the decoupled flavour direction orthogonal to
$\ell_{e\mu}$. However, since the corresponding doublets do not
participate in the leptogenesis dynamics, this asymmetry is irrelevant
for baryogenesis, and thus giving just the two entries corresponding
to $\ell_\tau$ and $\ell_{e\mu}$ in the matrices $A$ and $C$ above is
sufficient.

\subsubsection{Above the EW sphaleron 
equilibration temperature: the $\mathbf{B=0}$ regime.}
\label{sec:Beq0}

EW sphaleron processes take place at a rate per unit volume
$\Gamma/V\propto T^4\alpha_W^5\log(1/\alpha_W)$
\cite{ar98,bodeker98,ar97,be03}, and are expected to be in equilibrium
up to temperatures of about $\sim 10^{12}$~GeV.  Even for moderate
values of $\tan\beta$, the $b$ and $\tau$ Yukawa interactions are
still in equilibrium at this temperature. We will then consider here
the temperature regime $10^{12}-10^{13}\,$GeV in which supersymmetric
leptogenesis still occurs in the two flavour regime, but with EW
sphalerons switched off. The related chemical equilibrium condition
\eqref{eq:tEWmu} is then replaced by baryon number conservation. We set
$B=0$ as initial condition since a dynamical generation of the baryon
asymmetry is precisely the goal of leptogenesis.  This also implies
that $Y_{\Delta_\alpha}=-Y_{\Delta L_\alpha}$ and therefore the total
lepton asymmetry in the direction orthogonal to $\ell_{e\mu}$ vanishes
exactly. For this regime we obtain:
%{\bf agreed by all after corrections in v5}:
%
\begin{eqnarray}
\nonumber
A^\ell&=&\frac{1}{6\times 73327}\left(
\begin{array}{ccc}
-72807 &\ \,\, 1480&-5676\\
\ \, -77 &-50984&-4236 \\
\end{array}\right), \\ [5pt]  
C^{\tilde g}&=&\frac{1}{73327}\left(-130,\;-370,\;1419\right),\\ [5pt]
C^{\tilde H_u}&=&\frac{-1}{2 \times 73327}\left(9187,\;9226,\;686\right),\\ [5pt]
C^{\tilde H_d} &=& 
\frac{1}{6\times 73327}\left(1531,\;9998,\;-1482\right). 
\label{eq:ACT5}
\end{eqnarray}

\bigskip

\subsubsection{Above the temperature for $\mathbf\tau$ Yukawa  
equilibration: the one flavour regime.}

\medskip

For moderate values of $\tan\beta$, at temperatures above $T\sim
10^{13}\,$GeV we can set $h_\tau\to 0$.  This condition defines the
one flavour regime, in which only the dynamics of the lepton doublet
$\ell_1$ to which $N_1$ couples is important.  This allows us to set
to zero the two lepton density asymmetries in the direction orthogonal
to $\ell_1$: $Y_{\Delta L_{1_\perp}}= Y_{\Delta L'_{1_\perp}}=0$ and
we are left with just one non-vanishing lepton flavour asymmetry that
is $Y_{\Delta L_1}$.  On the other hand, approximate $b-\tau$ Yukawa
unification suggests that we should also set $h_b\to 0$, so that only
the $h_t$ Yukawa condition \eqref{eq:upquarks} remains.  Since quark
mixing between the second and third generation is, for small
$\tan\beta$, of the same order than the ratio $h_b/h_t \sim {\cal
  O}(10^{-2})$, we assume that $Q_2$ is also decoupled from Yukawa
interactions and thus we replace the condition $Q_2=Q_3$ by the
conservation of the corresponding charge $\chi_{Q_2}+\kappa_{Q_2}R_3$,
that yields the constraint \eqref{eq:chiralmuquarks2}.

Note that even if in this regime all lepton flavour effects can be
neglected~\cite{barbieri,flavour1,flavour2}, not all spectator effects
can~\cite{spectator2}. For example, QCD sphaleron equilibration is
maintained up to higher temperatures than the corresponding
electroweak processes because they have larger rates:
$\Gamma_{QCD}/\Gamma_{EW} \simeq 12 (\alpha_s/\alpha_W)^5$, and are
likely to be still in equilibrium at $T\sim
10^{13}$~GeV~\cite{mo97,mo92,be03}.  Since $Y_{\Delta_{\tilde N}}$
remains coupled to the lepton flavour dynamics because of its
contribution to ${\cal R}_{\rm tot}$ and ${R_3}_{\rm tot}$ we still
have matrices with two columns $(-Y_{\Delta_{L_1}},\;Y_{\Delta_{\tilde
    N}})$ even in the one flavour regime.  After imposing all the
relevant conservation conditions, including the vanishing of lepton
number in two flavour directions, we obtain:
\begin{eqnarray}
\nonumber
A^\ell=-\frac{1}{295}\left(49,\;4 \right)\,, &\quad&
C^{\tilde g}=\frac{1}{4\times 295}\left(-1,\;24 \right)\,, \\ [5pt] 
C^{\tilde H_u}= \frac{-7}{16\times 295}\left(49,\;4\right)\,, &\quad&
C^{\tilde H_d} = \frac{1}{6\times 295}\left(1,\;-24\right)\,.  
\label{eq:ACT6}
\end{eqnarray}
Before concluding this section, it is worth reminding that the
conversion of the $B-L$ asymmetry generated by the leptogenesis
dynamics into a baryon asymmetry should eventually be computed at a
temperature ${\cal O}(100\,{\rm GeV})$,  that is right before EW
sphaleron transitions are switched off. At this temperature
presumably all the supersymmetric particles have already decayed, and
the particle content of the thermal bath is then the same than in the SM
with two Higgs doublets. Depending if this temperature 
is higher or lower than the temperature $T_{\rm EWPT}$ of the EW  
phase transition, the conversion factors are~\cite{Laine:1999wv}:
\be
Y_{\Delta_B} = 
Y_{\Delta_{B-L}}\times 
\left\{\begin{array}{l} 
\frac{8}{23} \qquad T > T_{\rm EWPT} \\
\frac{10}{31} \qquad T < T_{\rm EWPT}\,. 
\end{array}\right.
\ee

\section{Numerical Results}
\label{sec:results}

As we have seen in the previous section, supersymmetric leptogenesis
is characterized by important qualitative differences with respect to
SM leptogenesis. These differences arise because in the NSE regime,
which is the relevant one to ensure that a sufficient amount of baryon
asymmetry can be generated, supersymmetric leptogenesis cannot be
treated by simply accounting for the new degrees of freedom in the
leptogenesis dynamics.  In this section we analyze the quantitative
relevance of the new effects we have been discussing.  They are all
related with washout effects that in the NSE regime are controlled by
the $3\times 4$ matrix ${\cal W}_{NSE}=A_{NSE}^\ell + C_{NSE}^{\tilde
  H_u} + C_{NSE}^{\tilde g}$, while in the SE regime the corresponding
matrix ${\cal W}_{SE}=A_{SE}^\ell + C_{SE}^{\tilde H_u}$ is $3\times
3$.  We can distinguish three types of effects:
\begin{enumerate}\itemsep 1pt
\item[(i)] The overall strength of the washout
  processes, that is mainly controlled by the diagonal entries of the
  $3\times 3 $ submatrix  ${\cal W}_{NSE}^{3\times 3}$, has slightly
  larger weights.

\item[(ii)] Lepton flavour mixing effects, that are controlled by the
  off diagonal entries in  ${\cal W}_{NSE}^{3\times 3}$ are sizeably larger in
  the NSE regime.

\item[(iii)] New contributions to the washout arise from mixing with
  $Y_{\Delta_{\tilde N}}$, given by the entries in the fourth column
  ${\cal W}_{NSE}^{\alpha 4}$.
\end{enumerate}
One can get an intuitive handle about the quantitative impact of these
effects on the final result for the baryon asymmetry, by comparing the
overall washout coefficients evaluated in the SE regime for the case
of the first generation Yukawa reactions out of equilibrium given in
\Eqn{eq:ACT3SE}, to the corresponding coefficients in the NSE regime
given in \Eqn{eq:ACT3NSE}.  We find, approximately independently of
the particular entry:
\begin{equation}
\left[{\cal W}_{NSE}^{3\times 3}-{\cal W}_{SE}\right]_{\alpha\beta} \approx 
-{\cal W}_{NSE}^{\alpha 4}\approx 
-0.01. 
\label{eq:WNSEWSE}
\end{equation}
For the (larger) diagonal entries this difference remains somewhat
below 10\%; for the (smaller) off-diagonal entries in ${\cal
  W}_{NSE}^{3\times 3}$ this represents an ${\cal O}(1)$ difference,
while the ${\cal W}_{NSE}^{\alpha 4}$ entries are specific of the NSE
regime.  We have found that in the regimes in which the washouts are
rather strong, and depending on the specific flavour configuration,
the first two effects can produce ${\cal O}(1)$ changes in the final
value of the baryon asymmetry. The last correction, although
potentially of a similar size, gives some effects on the evolution of
the flavour density asymmetries only for $z\lsim 1$, that is when the
$\tilde N$'s approach an equilibrium distribution. However, at $z >1$, 
$Y_{\Delta_{\tilde N}}$ gets exponentially suppressed implying    
that basically no effects are left in the final result.

\begin{figure}[t!]
 \includegraphics[width=\textwidth]{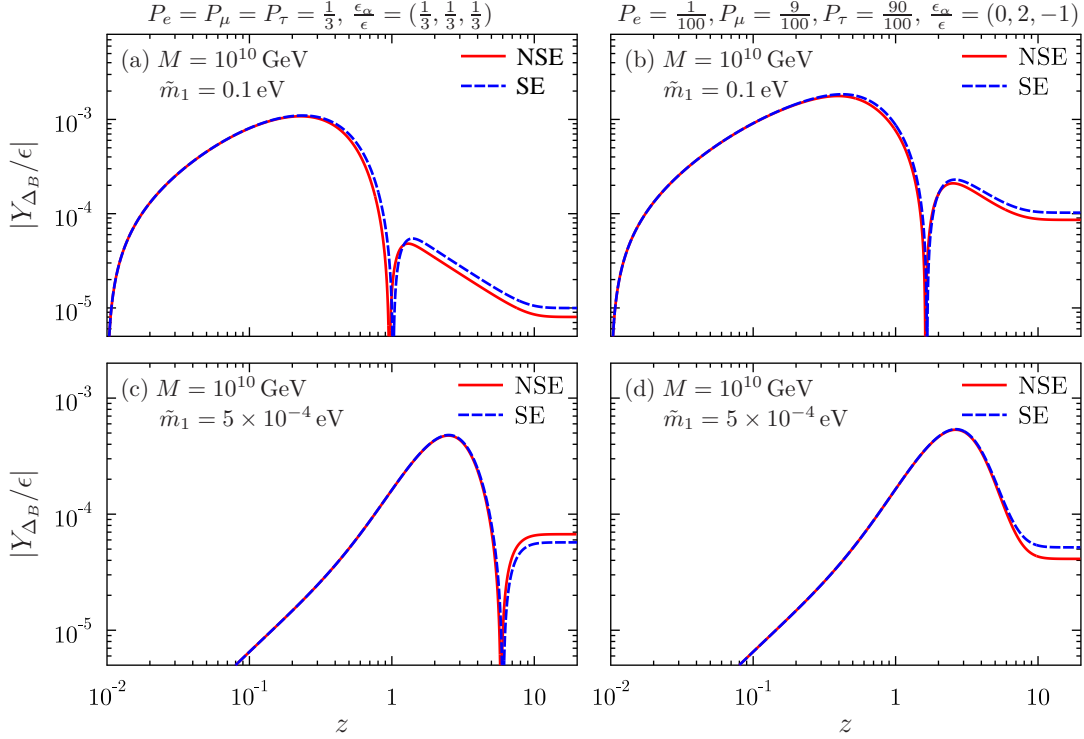}
 \caption[]{The baryon asymmetry normalized to the total CP asymmetry
   $|Y_{\Delta_B}/\epsilon|$ computed in the three flavour NSE regime with
   the $A$ and $C$ matrices in \eqref{eq:ACT3NSE} (solid red
   lines) confronted with what is obtained in the SE regime
   \eqref{eq:ACT3SE} (dashed blue lines).  The two upper panels (a)
   and (b) correspond to strong washouts with $\tilde m_1=0.1\,$eV, the
   two lower panels (c) and (d) to weak washouts with $\tilde
   m_1=5\times 10^{-4}\,$eV.  In the two panels (a) and (c) on the
   left a flavour equipartition configuration has been assumed, with
   flavoured parameters $\epsilon_\alpha/\epsilon=P_\alpha=1/3$.  In
   the two panels (b) and (d) on the right a strongly misaligned
   flavour configuration has been used, with $\epsilon_e=0$,
   $\epsilon_\mu=-2\epsilon_\tau=\epsilon$ and $P_e=1/100$,
   $P_\mu=9/100$ $P_\tau=90/100$.}
\label{fig:1}
\end{figure}

The results of our numerical analysis are summarized by the plots in
figures~\ref{fig:1} and \ref{fig:2}, that are obtained by integrating
numerically the complete set of Boltzmann equations given in the
Appendix. They include decays, inverse decays and scatterings with
top-quarks, and hold under the assumption that the $N$'s are
hierarchical and that the lepton asymmetries only result from the
decays of the lightest heavy singlet states $N\equiv N_{1}$.
However, in order to illustrate how the new effects described above
modify the structure of the equations, here we write much simpler
expressions in which only decays and inverse decays are included:
\begin{eqnarray}
\dot{Y}_{N} & = & -\left(\frac{Y_{N}}{Y_{N}^{eq}}-1\right)\gamma_{N},
\label{eq:YN}\\
\dot{Y}_{\tilde N_+} & = & 
-\left(\frac{Y_{\tilde N_+}}{Y_{\widetilde{N}}^{eq}}-2\right)
\gamma_{\widetilde{N}},
\label{eq:YNplus}\\
\dot{Y}_{\Delta_{\tilde N}} & = & 
-\frac{Y_{\Delta_{\tilde N}}}{Y_{\widetilde{N}}^{eq}}
\gamma_{\widetilde{N}}
-\frac{3}{2}\,\gamma_{\widetilde{N}}
\sum_a C_a^{\tilde g}\;\frac{Y_{\Delta_a}}{Y^{eq}_\ell}
 +\dots\,,
 \label{eq:YNminus}\\
\dot{Y}_{\Delta_{\alpha}} & = & -\epsilon_{\alpha}
\left[\left(\frac{Y_{N}}{Y_{N}^{eq}}-1\right)
\gamma_{N}+\left(\frac{Y_{\tilde N_+}}{Y_{\widetilde{N}}^{eq}}-2\right)
\gamma_{\widetilde{N}}\right] 
\nonumber 
 \\
 &&
+ \left(\gamma_{\widetilde{N}}^{\alpha}+\frac{1}{2}\gamma_{N}^{\alpha}\right)
\sum_a\left(A_{\alpha a}^\ell+ C_a^{\tilde H_u} 
+ C_a^{\tilde g}\right)\frac{Y_{\Delta_a}}{Y^{eq}_\ell}\,.
\label{eq:YDelta}
\end{eqnarray}
In the equations above the time derivative is defined as $\dot Y
\equiv zsH\, dY/dz$ with $s$ the entropy density, $z=M/T$ and $H\equiv
H(M)$ the expansion parameter at $T=M$. In \Eqn{eq:YNplus} we have
introduced the overall sneutrino abundance $Y_{\tilde N_+}=Y_{\tilde
  N}+ Y_{\tilde N^*}$, while $Y_{\Delta_{\tilde N}} \equiv Y_{\tilde
  N}- Y_{\tilde N^*}$ in \Eqn{eq:YNminus} is the sneutrino density
asymmetry that was already introduced in Section
\ref{sec:NSEconstraints}.  In the washout terms we have normalized the
charge densities $Y_{\Delta_a}=(Y_{\Delta_\alpha},\;Y_{\Delta_{\widetilde{N}}}) $ to the equilibrium density of a
fermion with one degree of freedom $Y_\ell$, and we refer to the
Appendix for detailed definitions of the reaction densities and
associated quantities.  In \Eqns{eq:YN}{eq:YDelta} we have also
neglected for simplicity all finite temperature effects.  Taking these
effects into account would imply for example that the CP asymmetry for
$\tilde N$ decays into fermions is different from the one for decays
into scalars, while we describe both CP asymmetries with
$\epsilon_\alpha$.  A few remarks regarding \Eqn{eq:YNminus} are in
order.  In the SE regime $\tilde g=0$ and thus it would seem that the
sneutrino density asymmetry $Y_{\Delta_{\tilde N}}$ vanishes.
However, this only happens for decays and inverse decays, and it is no
more true when additional terms related to scattering processes, that
are represented in the equation by the dots, are also included (see
ref.~\cite{Plumacher:1997ru} and the Appendix).  Therefore, also in
the SE regime $Y_{\tilde N}$ and $Y_{\tilde N^*}$ in general
differ. However, in this case recasting their equations in terms of
two equations for $Y_{\tilde N_+}$ and $Y_{\Delta_{\tilde N}}$ is just
a convenient parametrization. On the contrary, in the NSE regime this
is mandatory, because the sneutrinos carry a globally conserved
$R$-charge and $Y_{\Delta_{\tilde N}}$ is required to formulate
properly the corresponding conservation law. As we have seen, this
eventually results in $Y_{\Delta_{\tilde N}}$ contributing to the
expressions of the lepton flavour density asymmetries in terms of
slowly varying quantities.

%%%%%%%%%%%%%%%%%%%%%%%%%%%%%%%%%%%%%%%%%%%%%%%%%%%%%%%%%%%%%%%%%%%%%%%
% PANEL
%
%%%%%%%%%%%%%%%%%%%%%%%%%%%%%%%%%%%%%%%%%%%%%%%%%%%%%%%%%%%%%%%%%%%%%%%

The results for the final baryon asymmetry (normalized to the total CP
asymmetry $\epsilon=\sum_\alpha\epsilon_\alpha)$ are depicted in
Figure~\ref{fig:1}. We have fixed the value of the $N$ mass to
$M=10^{10}\,$GeV that is, we work in the three flavour regime in which
all the Yukawa couplings of the fermions of the first generation can
be set to zero and the results of Section \ref{sec:ACT3NSE} hold.  We
confront our results with the SE case discussed in
Section~\ref{sec:ACT3SE} in which the SE conditions $\tilde g=0$ and
$Y_{\Delta{\ell,\Delta H_{u}}}=2\,Y_{\Delta{\tilde\ell,\Delta\tilde
    H_{u}}}$ hold.  As it  is customary we parametrize the overall
strength of the washouts in terms of the effective 
mass~\cite{Fischler:1990gn,Plumacher:1996kc}
\begin{equation}
\label{eq:tm}
\tilde m_1=\sum_{\alpha}\frac{|\lambda_{1\alpha}|^2 v_u^2}{M}
\equiv \sum_{\alpha}\tilde m_{1\alpha}\equiv 
\sum_{\alpha}P_{\alpha}\tilde m_{1}
\; , 
\end{equation}
where $v_u$ is the vacuum expectation value of the up-type Higgs
doublet, $v_u=v\, \sin\beta $ ($v$=174 GeV).\footnote{ Once the set of
  Yukawa conditions used to derive the $A$ and $C$ matrices is
  established, the dominant dependence on $\tan\beta$ in the numerical
  results arises via $v_u$ in \Eqn{eq:tm} and therefore it is very
  mild.}  For the single lepton flavours, the corresponding parameters
are defined in terms of the tree level $N$ and $\tilde N$ decays
branching ratios
$P_\alpha=|\lambda_{1\alpha}|^2/(\lambda\lambda^\dagger)_{11}$
according to:
\begin{equation}
\label{eq:tmalpha}
\tilde m_{1\alpha}\equiv P_{\alpha}\tilde m_{1}, 
\qquad \qquad 
\sum_\alpha P_\alpha=1. 
\end{equation}
In Figure~\ref{fig:1} the solid red lines represent the results
obtained with the NSE $A$ and $C$ matrices in Eq.\eqref{eq:ACT3NSE},
while the dashed blue lines give the corresponding results obtained
assuming SE and using $A$ and $C$ in Eq.\eqref{eq:ACT3SE}.  In the two
upper panels (a) and (b) we give the results obtained assuming a
particularly strong washout regime corresponding to $\tilde
m_1=0.1\,$eV, while the two lower panels (c) and (d) correspond to a
weak washout regime with $\tilde m_1=5\times 10^{-4}\,$eV.  In the
attempt to disentangle the differences between NSE and SE due to the
changes in the overall washout strength [effects of type (i)] from
those due to the changes in the flavour mixing pattern [effects of
type (ii)] we have assumed for the two panels (a) and (c) on the left
a flavour equipartition configuration
[$\frac{\epsilon_\alpha}{\epsilon}= \frac{1}{3}$;
$P_\alpha=\frac{1}{3}$] while for the two panels (b) and (d) on the
right we have assumed a strongly misaligned flavour configuration
[$\frac{\epsilon_\alpha}{\epsilon}=(0,\;2,\;-1)$; $P_\alpha=(
\frac{1}{100},\; \frac{9}{100},\; \frac{90}{100})$].  In the first
case the differences in the flavour mixing patterns produce irrelevant
effects, and the changes that can be seen are essentially due to the
different flavour diagonal washouts.  In the second case the
differences in the patterns of flavour mixing also contribute.  As it
is apparent from figure~\ref{fig:1}, the numerical differences between
the NSE and SE cases remain typically at the ${\cal O}(1)$ level.  As
regards the effects of type (iii) that are related to
$Y_{\Delta_{\tilde N}}$, we have verified that in all the plots in
figure~\ref{fig:1} any visible difference would appear only around
$z\lsim 1$, leaving the final results unchanged.  However, for
completeness, we present in figure~\ref{fig:2} the detailed evolution
of the single flavour charge densities $Y_{\Delta_\alpha}$ and also of
$Y_{\Delta_{\tilde N}}$ corresponding to the flavour/washout
configuration of figure~\ref{fig:1}(b).  As in the previous figure,
the NSE results are depicted with solid lines [from up to down and
around $z=0.5$: $Y_{\Delta_\mu}$ (blue), $Y_{\Delta_\tau}$ (magenta),
$Y_{\Delta_{\tilde N}}$ (black), $Y_{\Delta_e}$ (red)] and the SE
results are depicted with dashed lines.  The thicker portion of each
line corresponds to positive values of the corresponding asymmetry,
while the thinner portion to negative values.  A couple of interesting
features can be identified: firstly, the largest differences occur for
$Y_{\Delta_e}$ and this is because given that $\epsilon_e=0$, this
asymmetry evolves solely because of flavour mixing effects; secondly,
it can be seen how $Y_{\Delta_{\tilde N}}$ gets strongly suppressed
around the time when the flavour asymmetries `freeze out' ($z\sim 5$),
and this explains the irrelevance of its effects on the final
result. One can also notice that curiously in the NSE regime
$Y_{\Delta_{\tilde N}}$ changes sign twice, while only once in SE.
This is, of course, due to the related mixing effects that are absent
in the SE case.

%%%%%%%%%%%%%%%%%%%%%%%%%%%%%%%%%%%%%%%%%%%%%%%%%%%%%%%%%%%%%%%%%%%%%%%
% PANEL
\begin{figure}[t!]
 \includegraphics[width=\textwidth]{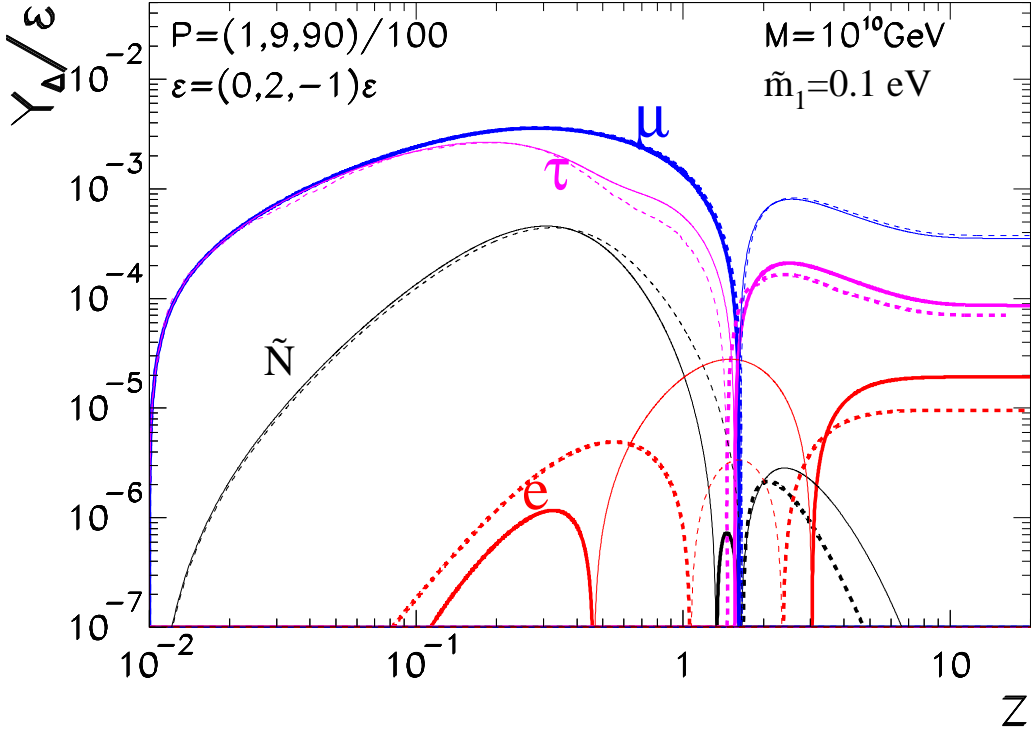}
 \caption[]{ Evolution of the normalized single density asymmetries
   $|Y_{\Delta_\alpha}/\epsilon|$ and $|Y_{\Delta_{\tilde
       N}}/\epsilon|$ for the flavour/washout configuration of
   figure~\ref{fig:1}(b). From up to down and around $z=0.5$:
   $Y_{\Delta_\mu}$ (blue), $Y_{\Delta_\tau}$ (magenta),
   $Y_{\Delta_{\tilde N}}$ (black) and $Y_{\Delta_e}$ (red).  Solid
   continuous lines give the results for the NSE regime, with the
   thicker portion of each line corresponding to positive values of
   the asymmetry, and the thinner portion to negative values.  The SE
   results are depicted with dashed lines and with the same
   color/flavour correspondence.}
\label{fig:2}
\end{figure} 
%
%%%%%%%%%%%%%%%%%%%%%%%%%%%%%%%%%%%%%%%%%%%%%%%%%%%%%%%%%%%%%%%%%%%%%%%

\section{Discussion and Conclusions} 
\label{sec:conclusions}

Motivated by all the recent advancements in leptogenesis studies, we
have revisited the theory of supersymmetric leptogenesis in the
attempt to put it on a more firm theoretical ground.  By digging in
some depth into the analysis of the various conditions that constrain
the density asymmetries of the different particles in the thermal
bath, we have found that important qualitative differences exist with
respect to SM leptogenesis.  We have first clarified the reasons why,
even if the specific constraining conditions are different for the
different temperature ranges in which leptogenesis can occur, the
whole set of particle density asymmetries can be always expressed in
terms of the same number of independent quantities.  This fact is
explained in the following way: whenever it happens that by raising
the temperature a chemical equilibrium condition ceases to hold
because the particle reaction enforcing it goes out of equilibrium,
then one Lagrangian parameter related to this out-of-equilibrium
reaction can be set to zero.  This generically results in a new global
symmetry, and the associated conservation law then enforces a new
condition for the particle number densities that replaces the one from
chemical equilibrium. Following this scheme, we have identified new
symmetries that are specific of the supersymmetric leptogenesis case.
New chemical potentials associated to the superpartners of the  SM
particles are also a generic feature of supersymmetric leptogenesis,
and a quite crucial example is provided by the gauginos.  At the
typical high temperatures relevant for leptogenesis, chirality
flipping transitions for the gauginos are completely out of thermal
equilibrium and basically do not occur. Gauginos thus develop a
chemical potential that can be thought of as the difference between
the number of L- and R-handed states. Three main consequences follow
from this. The first one is that particle and sparticle chemical
potentials are different during leptogenesis.  This is because only at
relatively low temperatures their chemical potentials are equilibrated
by scatterings. However, the rates of these scatterings vanish in the
limit of vanishing gaugino masses, and  are completely negligible
at the high temperatures relevant for leptogenesis.  This modifies the
weights of the scalars and fermions-related washouts.  The second one
is that by setting to zero the gaugino mass, a new $R$-symmetry
arises. While this symmetry is anomalous, a suitable anomaly free
combination can be constructed and this corresponds to an exactly
conserved global charge.  The related conservation constraint, that is
specific to supersymmetry, modifies the $A$ and $C$ matrices that
describe the mixing between lepton flavours induced by the EW
sphalerons.  The third one is that since sneutrinos carry a unit
charge under the  $R$ symmetry, their number density
asymmetry enters the corresponding conservation law. The result
is that through the $A$ and $C$  matrices the lepton flavour
asymmetries are now related to four independent quantities rather than
the usual three flavour charges of the SM case, being the sneutrino
density asymmetry the fourth one.

While we believe that the results presented in this paper are quite
interesting from the theoretical point of view, by comparing the
baryon asymmetry yield of leptogenesis within our framework with the
results obtained by neglecting all the new effects, we have concluded
that quantitatively we are dealing with ${\cal O}(1)$ corrections.
This is hardly surprising since, quite in general, sizable changes in
the baryon asymmetry generated through leptogenesis are related with
the identification of new sources of CP violation or by spoiling some
exact cancellation.  While this is the case e.g. for flavoured
leptogenesis~\cite{barbieri,flavour1,flavour2} that requires the
inclusion of the flavour CP asymmetries $\epsilon_\alpha$, or for soft
leptogenesis~\cite{soft1,soft2,soft3} where new CP violating phases
play a crucial role and thermal effects spoil an exact
zero-temperature cancellation between the CP asymmetries for decays
into scalars and fermions, in our case there are no changes in the
amount of CP violation that drives leptogenesis nor cancellations that
could be spoiled.

\acknowledgments 

This work is supported by  USA-NSF grant PHY-0653342 and by 
Spanish  grants from MICINN 2007-66665-C02-01, the 
INFN-MICINN agreement program ACI2009-1038,  consolider-ingenio 2010 
program  CSD-2008-0037 and by CUR Generalitat de Catalunya grant 2009SGR502.
We thank Yosef Nir for his comments on a preliminary version 
of the paper. 

\newpage
\appendix
\section{Boltzmann equations for supersymmetric leptogenesis}
% Our numerical results have been derived with 

The Boltzmann equations that we have used to derive the numerical results read:
\begin{eqnarray}
\dot{Y}_{N} & = & 
-\left(\frac{Y_{N}}{Y_{N}^{eq}}-1\right)
\left(\gamma_{N}+4\gamma_{t}^{(0)}+4\gamma_{t}^{(1)}+4\gamma_{t}^{(2)}
+2\gamma_{t}^{(3)}+4\gamma_{t}^{(4)}\right),
\label{eq:YN_full} \\ 
\dot{Y}_{\tilde N_+} & = & 
-\left(\frac{Y_{\tilde N_+}}{Y_{\widetilde{N}}^{eq}}-2\right)
\left(\gamma_{\widetilde{N}}+3\gamma_{22}
+2\gamma_{t}^{(5)}+2\gamma_{t}^{(6)}+2\gamma_{t}^{(7)}+\gamma_{t}^{(8)}+2\gamma_{t}^{(9)}\right)
%\nonumber \\
%&  & +\frac{Y_{\Delta_{\tilde N}}}{Y_{\widetilde{N}}^{eq}}
%\left[\frac{1}{2}\frac{Y_{\Delta\widetilde{\ell}}}{Y_{\widetilde{\ell}}^{eq}}
%\left(\gamma_{22}-\gamma_{t}^{(8)}\right)+\frac{Y_{\Delta\ell}}{Y_{\ell}^{eq}}\gamma_{t}^{(5)}
%+\frac{Y_{\Delta\widetilde{H}_{u}}}{Y_{\widetilde{H}_{u}}^{eq}}\left(\gamma_{22}+\gamma_{t}^{(6)}\right)\right.
%\nonumber \\
%&  & \left.+\frac{1}{2}\frac{Y_{\Delta H_{u}}}{Y_{H_{u}}^{eq}}
%\left(\gamma_{22}+\gamma_{t}^{(6)}+\gamma_{t_{i}}^{(7)}-\gamma_{t_{i}}^{(9)}\right)\right]
\; ,
\label{eq:YNplus_full} \\
\dot{Y}_{\Delta_{\tilde N}} & = & 
-\frac{Y_{\Delta_{\tilde N}}}{Y_{\widetilde{N}}^{eq}}
\left(\gamma_{\widetilde{N}}
+3\gamma_{22}+2\gamma_{t}^{(5)}
+2\gamma_{t}^{(6)}+2\gamma_{t}^{(7)}+\gamma_{t}^{(8)}+2\gamma_{t}^{(9)}\right)
+\frac{1}{2}\left(\frac{Y_{\Delta\ell}}{Y_{\ell}^{eq}}
-\frac{Y_{\Delta\widetilde{\ell}}}{Y_{\widetilde{\ell}}^{eq}}\right)\gamma_{\widetilde{N}}
\nonumber \\  &  & 
+\frac{Y_{\Delta\ell}}{Y_{\ell}^{eq}}\left(\frac{Y_{\tilde N_+}}{Y_{\widetilde{N}}^{eq}}\gamma_{t}^{(5)}
+2\gamma_{t}^{(6)}+2\gamma_{t}^{(7)}\right)
% \nonumber \\ &  & 
+\frac{Y_{\Delta\widetilde{\ell}}}{Y_{\widetilde{\ell}}^{eq}}
\left[\left(2
+\frac{1}{2}\frac{Y_{\tilde N_+}}{Y_{\widetilde{N}^{eq}}}\right)\gamma_{22}
-\frac{1}{2}\frac{Y_{\tilde N_+}}{Y_{\widetilde{N}}^{eq}}\gamma_{t}^{(8)}-2\gamma_{t}^{(9)}\right]
\nonumber \\
&  & +2\frac{Y_{\Delta\widetilde{H}_{u}}}{Y_{\widetilde{H}_{u}}^{eq}}
\left[\left(2+\frac{1}{2}\frac{Y_{\tilde N_+}}{Y_{\widetilde{N}}^{eq}}\right)\gamma_{22}
+\gamma_{t}^{(5)}+\frac{1}{2}\frac{Y_{\tilde N_+}}{Y_{\widetilde{N}}^{eq}}\gamma_{t}^{(6)}+\gamma_{t}^{(7)}\right]
 \nonumber \\&  & 
+\frac{Y_{\Delta H_{u}}}{Y_{H_{u}}^{eq}}
\left[\!\left(2+\frac{1}{2}\frac{Y_{\tilde N_+}}{Y_{\widetilde{N}}^{eq}}\right)\!\gamma_{22}\!
+2\gamma_{t}^{(5)}\!
% \right. \nonumber \\ &  &  \left.
+\left(\!1+\frac{1}{2}\frac{Y_{\tilde N_+}}{Y_{\widetilde{N}}^{eq}}\right)\!
\left(\!\gamma_{t}^{(6)}\!+\gamma_{t}^{(7)}\right)\!-\gamma_{t}^{(8)}\!-2\gamma_{t}^{(9)}\!\right],
\label{eq:YNminus_full}  \\
-\dot{Y}_{\Delta_{\alpha}} & = & \epsilon_{\alpha}
\left(\frac{Y_{N}}{Y_{N}^{eq}}-1\right)
\left(\gamma_{N}+4\gamma_{t}^{(0)}+4\gamma_{t}^{(1)}
+4\gamma_{t}^{(2)}+2\gamma_{t}^{(3)}+4\gamma_{t}^{(4)}\right)
\nonumber \\
&  & +\epsilon_{\alpha}\left(\frac{Y_{\tilde N_+}}{Y_{\widetilde{N}}^{eq}}-2\right)
\left(\gamma_{\widetilde{N}} 
+3\gamma_{22}+2\gamma_{t}^{(5)}+2\gamma_{t}^{(6)}
+2\gamma_{t}^{(7)}+\gamma_{t}^{(8)}+2\gamma_{t}^{(9)}\right)
\nonumber \\
&  & -\frac{1}{2}\left(\frac{Y_{\Delta\ell_{\alpha}}}{Y_{\ell}^{eq}}
+\frac{Y_{\Delta\widetilde{\ell}_{\alpha}}}{Y_{\widetilde{\ell}}^{eq}}
+\frac{Y_{\Delta H_{u}}}{Y_{H}^{eq}}
+\frac{Y_{\Delta\widetilde{H}_{u}}}{Y_{\widetilde{H}}^{eq}}\right)
\left(\gamma_{\widetilde{N}}^{\alpha}+\frac{1}{2}\gamma_{N}^{\alpha}\right)
\nonumber \\
&  & -\left(\frac{Y_{\Delta\widetilde{\ell}_{\alpha}}}{Y_{\widetilde{\ell}}^{eq}}
+2\frac{Y_{\Delta\widetilde{H}_{u}}}{Y_{\widetilde{H}_{u}}^{eq}}
-\frac{Y_{\Delta H_{u}}}{Y_{H_{u}}^{eq}}\right)
\left(\frac{1}{2}\frac{Y_{\tilde N_+}}{Y_{\widetilde{N}}^{eq}}
+2\right)\gamma_{22}^{\alpha}
\nonumber \\
&  & -\frac{Y_{\Delta\ell_{\alpha}}}{Y_{\ell}^{eq}}\left(\frac{Y_{N}}{Y_{N}^{eq}}
\gamma_{t}^{(3)\alpha}+2\gamma_{t}^{(4)\alpha}
+\frac{Y_{\tilde N_+}}{Y_{\widetilde{N}}^{eq}}\gamma_{t}^{(5)\alpha}
+2\gamma_{t}^{(6)\alpha}+2\gamma_{t}^{(7)\alpha}\right)
\nonumber \\
&  & -\frac{Y_{\Delta\widetilde{\ell}_{\alpha}}}{Y_{\widetilde{\ell}}^{eq}}
\left(2\frac{Y_{N}}{Y_{N}^{eq}}\gamma_{t}^{(0)\alpha}
+2\gamma_{t}^{(1)\alpha}+2\gamma_{t}^{(2)\alpha}
+\frac{1}{2}\frac{Y_{\tilde N_+}}{Y_{\widetilde{N}}^{eq}}\gamma_{t}^{(8)\alpha}
+2\gamma_{t}^{(9)\alpha}\right)
\nonumber \\
&  & -\frac{Y_{\Delta H_{u}}}{Y_{H_{u}}^{eq}}
\left[\frac{1}{2}\left(\frac{Y_{\tilde N_+}}{Y_{\widetilde{N}}^{eq}}-2\right)
\left(-\gamma_{t}^{(6)\alpha}+\gamma_{t}^{(7)\alpha}\right)
+\gamma_{t}^{(8)\alpha}+\left(1+\frac{1}{2}\frac{Y_{\tilde N_+}}{Y_{\widetilde{N}}^{eq}}\right)
\gamma_{t}^{(9)\alpha}\right]
\nonumber \\
&  & -\frac{Y_{\Delta H_{u}}}{Y_{H_{u}}^{eq}}
\left[\left(\frac{Y_{N}}{Y_{N}^{eq}}-1\right)\gamma_{t}^{(1)\alpha}
-\frac{Y_{N}}{Y_{N}^{eq}}\gamma_{t}^{(2)\alpha}+\gamma_{t}^{(3)\alpha}
+\left(1+\frac{Y_{N}}{Y_{N}^{eq}}\right)\gamma_{t}^{(4)\alpha}\right]
\nonumber \\
 &  & -2\frac{Y_{\Delta\widetilde{H}_{u}}}{Y_{\widetilde{H}_{u}}^{eq}}
\left(\gamma_{t}^{(0)\alpha}+\gamma_{t}^{(1)\alpha}
+\frac{Y_{N}}{Y_{N}^{eq}}\gamma_{t}^{(2)\alpha}+\gamma_{t}^{(5)\alpha}
+\gamma_{t}^{(7)\alpha}+\frac{1}{2}\frac{Y_{\tilde N_+}}{Y_{\widetilde{N}}^{eq}}\gamma_{t}^{(6)\alpha}\right)
\nonumber \\
&  & +\frac{Y_{\Delta_{\tilde N}}}{Y_{\widetilde{N}}^{eq}}
\left(2\gamma_{t}^{(5)\alpha}+2\gamma_{t}^{(6)\alpha}
+2\gamma_{t}^{(7)\alpha}-\gamma_{t}^{(8)\alpha}-2\gamma_{t}^{(9)\alpha}\right).
\label{eq:YDelta_full} 
\end{eqnarray}
Besides the decays and inverse decays included in
\Eqns{eq:YN}{eq:YDelta}, these equations also include scatterings with
the top-quark, both in the washout and in the CP asymmetries.  In the
limit in which neutrinos are sufficiently hierarchical, as we are
assuming here, the CP asymmetries in scatterings with top quarks 
are the same than the
CP asymmetries in decays~\cite{flavour3,CPscatt} and then can be
easily included.  We have not included gauge boson scatterings nor the
corresponding CP asymmetries.  We have neglected three body decays
since their contribution is phase space suppressed, and also because
we have found that in ref.~\cite{Plumacher:1997ru}, on which our
equations are based, some diagrams related to these processes have
been overlooked.  $\Delta L=2$ scatterings mediated by off-shell
singlet neutrinos are also left out since in the temperature range
$T\sim 10^{10}\,$GeV in which our results are obtained they are
completely irrelevant.  We have also neglected all finite temperature
effects except for the Higgs thermal mass that is kept to regulate the
infrared divergences in scatterings with the top-quark.

We define the equilibrium densities per degree of freedom normalized
to the entropy density as:
\begin{eqnarray}
Y_{\ell}^{eq} & = & \frac{1}{2}Y_{\widetilde{\ell}}^{eq}
=\frac{1}{2}Y_{H_{u}}^{eq}=Y_{\widetilde{H}_{u}}^{eq}
=\frac{15}{8\pi^{2}g_{*}},
\label{eq:nor_density}
\end{eqnarray}
with the MSSM number of effective degrees of freedom $g_{*}=228.75$.
The number density asymmetries are defined  according to 
$Y_{\Delta\ell}=  Y_{\ell}- Y_{\bar\ell}$.  
Density asymmetries and reaction densities without an  
index $\alpha$ are understood to be summed
over flavours: $Y_{\Delta\ell}=\sum_{\alpha}Y_{\Delta \ell_{\alpha}}$ 
and $\gamma_N=\sum_{\alpha}\gamma_N^{\alpha}$. 
In \Eqns{eq:YN}{eq:YDelta} the density asymmetries for scalars 
$Y_{\Delta\widetilde{\ell}_{\alpha}}$
and $Y_{\Delta H_{u}}$ can be expressed in terms of 
the corresponding  asymmetries for fermions 
$Y_{\Delta\ell_{\alpha}}$ and
$Y_{\Delta\widetilde{H}_{u}}$ according to:
\begin{eqnarray}
\frac{Y_{\Delta\widetilde{\ell}_{\alpha}}}{Y_{\widetilde{\ell}}^{eq}} 
& = & \frac{Y_{\Delta\ell_{\alpha}}}{Y_{\ell}^{eq}}
+\frac{Y_{\Delta\widetilde{g}}}{Y_{\widetilde{g}}^{eq}},\\
\frac{Y_{\Delta H_{u}}}{Y_{H_{u}}^{eq}} 
& = & \frac{Y_{\Delta\widetilde{H}_{u}}}{Y_{\widetilde{H}_{u}}^{eq}}
+\frac{Y_{\Delta\widetilde{g}}}{Y_{\widetilde{g}}^{eq}},
\end{eqnarray}
where $Y_{\widetilde{g}}^{eq}=Y_{\widetilde{H}_{u}}^{eq}=Y_{\ell}^{eq}$.
In the SE regime,  $Y_{\Delta_{\widetilde g}}=0$ and thus 
$\frac{Y_{\Delta\widetilde{\ell}_{\alpha}}}{Y_{\widetilde{\ell}}^{eq}} 
 =  \frac{Y_{\Delta\ell_{\alpha}}}{Y_{\ell}^{eq}}$ and
$\frac{Y_{\Delta H_{u}}}{Y_{H_{u}}^{eq}} 
 =  \frac{Y_{\Delta\widetilde{H}_{u}}}{Y_{\widetilde{H}_{u}}^{eq}}$
follow.
The reaction densities entering \Eqns{eq:YN_full}{eq:YDelta_full} are:
\begin{eqnarray}
\gamma^{\alpha}_N &\equiv& 
\gamma\left(N \leftrightarrow \widetilde{\ell}_{\alpha} 
\widetilde{H}_u \right)
+\gamma\left(N \leftrightarrow \widetilde{\ell}_{\alpha}^* 
\overline{\widetilde{H}_u} \right)
+\gamma\left(N \leftrightarrow \ell_{\alpha} H_u \right)
+\gamma\left(N \leftrightarrow \overline{\ell}_{\alpha} H_u^* \right) ,
\nonumber \\
\gamma^{\alpha}_{\widetilde N} &\equiv&
\gamma \left(\widetilde{N} \leftrightarrow
\widetilde{\ell}_{\alpha} H_u \right)
+\gamma \left(\widetilde{N} \leftrightarrow
\overline{\ell_{\alpha}} \overline{\widetilde{H}_u} \right)
=\gamma \left(\widetilde{N}^* \leftrightarrow
\widetilde{\ell}_{\alpha}^* H_u^* \right)
+\gamma \left(\widetilde{N}^* \leftrightarrow
\ell_{\alpha} \widetilde{H}_u \right) ,
\nonumber \\
\gamma_{22}^{\alpha} 
& \equiv & \gamma\left(\widetilde{N}\widetilde{\ell}_{\alpha}\leftrightarrow\widetilde{Q}\widetilde{u}^*\right)
=\gamma\left(\widetilde{N}\widetilde{Q}^{*}\leftrightarrow\widetilde{\ell}_{\alpha}^{*}\widetilde{u}^*\right)
=\gamma\left(\widetilde{N}\widetilde{u}\leftrightarrow\widetilde{\ell}_{\alpha}^{*}\widetilde{Q}\right),
\nonumber \\
\gamma_{t}^{(0)\alpha} 
& \equiv & \gamma\left(N\widetilde{\ell}_{\alpha}\leftrightarrow Q\widetilde{u}^*\right)
=\gamma\left(N\widetilde{\ell}_{\alpha}\leftrightarrow\widetilde{Q}\overline{u}\right),
\nonumber \\
\gamma_{t}^{(1)\alpha} 
& \equiv & \gamma\left(N\overline{Q}\leftrightarrow\widetilde{\ell}_{\alpha}^{*}\widetilde{u}^*\right)
=\gamma\left(Nu\leftrightarrow\widetilde{\ell}_{\alpha}^{*}\widetilde{Q}\right),
\nonumber \\
\gamma_{t}^{(2)\alpha} 
& \equiv & \gamma\left(N\widetilde{u}\leftrightarrow\widetilde{\ell}_{\alpha}^{*}Q\right)
=\gamma\left(N\widetilde{Q}^{*}\leftrightarrow\widetilde{\ell}_{\alpha}^{*}\overline{u}\right),
\nonumber \\
\gamma_{t}^{(3)\alpha} 
& \equiv & \gamma\left(N\ell_{\alpha}\leftrightarrow Q\overline{u}\right),
\nonumber 
\end{eqnarray}
\begin{eqnarray}
\gamma_{t}^{(4)\alpha} 
& \equiv & \gamma\left(Nu\leftrightarrow\overline{\ell_{\alpha}}Q\right)
=\gamma\left(N\overline{Q}\leftrightarrow\overline{\ell_{\alpha}}\overline{u}\right),
\nonumber \\
\gamma_{t}^{(5)\alpha} 
& \equiv & \gamma\left(\widetilde{N}\ell_{\alpha}\leftrightarrow Q\widetilde{u}^*\right)
=\gamma\left(\widetilde{N}\ell_{\alpha}\leftrightarrow\widetilde{Q}\overline{u}\right),
\nonumber \\
\gamma_{t}^{(6)\alpha} 
& \equiv & \gamma\left(\widetilde{N}\widetilde{u}\leftrightarrow\overline{\ell_{\alpha}}Q\right)
=\gamma\left(\widetilde{N}\widetilde{Q}^{*}\leftrightarrow\overline{\ell_{\alpha}}\overline{u}\right),
\nonumber \\
\gamma_{t}^{(7)\alpha} 
& \equiv & \gamma\left(\widetilde{N}\overline{Q}\leftrightarrow\overline{\ell_{\alpha}}\widetilde{u}^*\right)
=\gamma\left(\widetilde{N}u\leftrightarrow\overline{\ell_{\alpha}}\widetilde{Q}\right),
\nonumber \\
\gamma_{t}^{(8)\alpha} 
& \equiv & \gamma\left(\widetilde{N}\widetilde{\ell}_{\alpha}^{*}\leftrightarrow\overline{Q}u\right),
\nonumber \\
\gamma_{t}^{(9)\alpha} 
& \equiv & \gamma\left(\widetilde{N}Q\leftrightarrow\widetilde{\ell}_{\alpha}u\right)
=\gamma\left(\widetilde{N}\overline{u}\leftrightarrow\widetilde{\ell}_{\alpha}\overline{Q}\right).
\end{eqnarray}
The reduced cross sections for the processes listed above can be found 
in ref.~\cite{Plumacher:1997ru}.

\vspace{2truecm}

\end{document}